\documentclass[a4paper,11pt]{article}
\usepackage{jcappub} 
\usepackage{slashed}
\usepackage{subfloat}  
\usepackage{comment}
\usepackage{subcaption}
\usepackage{tikz}                   
\usepackage{float}
\usepackage{booktabs} 
\usepackage[table]{xcolor} 
\usepackage{booktabs}
\usepackage{caption}
\usepackage{float}
\usepackage{graphicx}
\usepackage{subcaption}
\usepackage{float}
\usepackage{booktabs}
\usepackage{mathrsfs}
\usepackage[T1]{fontenc} 
\definecolor{lightgray}{RGB}{245, 245, 245}
\definecolor{headerblue}{RGB}{220, 230, 250}
\definecolor{acsblue}{RGB}{17,76,139}
\definecolor{shadecolor}{RGB}{255,241,204}
\newcommand{{\ri}}{{\rm{i}}}
\newcommand{{\Psibar}}{{\bar{\Psi}}}

\newcommand{\bra}{\begin{array}}
	\newcommand{\era}{\end{array}}
\newcommand{\beq}{\begin{equation}}
	\newcommand{\eeq}{\end{equation}}
\newcommand{\beqar}{\begin{eqnarray}}
	\newcommand{\eeqar}{\end{eqnarray}}

\newcommand{\be}{\begin{equation}}
	\newcommand{\ee}{\end{equation}}
\newcommand{\bea}{\begin{eqnarray}}
	\newcommand{\eea}{\end{eqnarray}}
\newcommand{\bd}{\begin{displaymath}}
	\newcommand{\ed}{\end{displaymath}}

\title{\huge\boldmath Ergosphere Dynamics and Rotational Energy Extraction in Bumblebee Kerr-Newman-AdS Black Holes}

\author[a,b,1]{H. Hassanabadi, \note{Corresponding author.}}
\author[c]{A. Guvendi,}
\author[d]{F. Kafikang,}
\author[e]{T. Sathiyaraj,}
\author[a,f]{S.Zare}

\affiliation[a]{Department of Physics, University of Hradec Kr$\acute{a}$lov$\acute{e}$, Rokitansk$\acute{e}$ho 62, 500 03 Hradec Kr$\acute{a}$lov$\acute{e}$, Czechia}
\affiliation[b]{Department of Physics and Electronics, Khazar University, 41 Mahsati Str, 1096 Baku, Azerbaijan}
\affiliation[c]{Department of Basic Sciences, Erzurum Technical University, 25050, Erzurum, Türkiye}
\affiliation[d]{Center for Theoretical Physics, Khazar University, 41 Mehseti Street, Baku, AZ-1096, Azerbaijan}
\affiliation[e]{Institute of Actuarial Science and Data Analytics, UCSI University, Cheras 56000, Kuala Lumpur, Malaysia}
\affiliation[f]{Helsinki Institute of Physics, University of Helsinki, P.O. Box 64, FI-00014, Helsinki, Finland}

\emailAdd{hha1349@gmail.com}
\emailAdd{abdullah.guvendi@erzurum.edu.tr}
\emailAdd{f.kafi19@yahoo.com}
\emailAdd{sathiyaraj133@gmail.com}
\emailAdd{soroushzrg@gmail.com}

\abstract{We present a comprehensive analysis of the thermodynamic and optical properties of the Bumblebee Kerr-Newman-Anti-de Sitter (AdS) black hole, a rotating and charged configuration arising in Lorentz symmetry-violating (LSV) gravity. The influence of the black hole parameters on the horizon structure, thermodynamic stability, and geometric deformation of spacetime is systematically investigated. Explicit expressions for the Hawking temperature, entropy, and heat capacity are derived, revealing the formation of black hole remnants and extended stability phases induced by Lorentz symmetry-violating (LSV) effects. The sparsity of Hawking radiation is quantified, showing that Lorentz violation suppresses the continuum limit and produces a more discrete, less thermal emission spectrum. A detailed analysis of null geodesics is performed to determine the photon region and shadow morphology, indicating that increasing $l$ and $Q$ compresses and distorts the shadow boundary, while rotation diminishes its overall size. The ergosphere geometry is analyzed in detail, showing that increases in $a$, $l$, and $Q$ not only enlarge and distort the ergoregion but also intensify frame-dragging, thereby maximizing the efficiency of energy extraction via the Penrose process. These results reveal clear and potentially observable deviations from standard Kerr-Newman-AdS predictions, providing a powerful new avenue to probe Lorentz symmetry breaking and test the fundamental structure of gravity in extreme strong-field regimes.}

\makeatletter
\gdef\@fpheader{}
\makeatother

\begin{document}
\maketitle
\flushbottom

\section{Introduction}\label{sec:intro}

Black holes (BHs) represent some of the most fascinating and extreme astrophysical objects in the universe, serving as natural laboratories for investigating strong gravitational fields, high-energy physics, and the fundamental structure of spacetime \cite{a,aa-8,a-1,a-2,aa-5,aa-6,aa-7}. They occupy a unique position at the intersection of astrophysics, cosmology, and fundamental physics, providing critical insights into phenomena such as gravitational lensing, accretion dynamics, quasiperiodic oscillations, Hawking radiation, and the tidal disruption of orbiting stars \cite{a-1,a-2,aa-6,aa-8}. Beyond the classical framework of general relativity (GR), BHs provide an unparalleled arena for testing modified gravity theories \cite{a-3,a-4}, probing quantum gravity proposals \cite{aa-9,aa-10}, and exploring higher-order corrections to GR on both microscopic and cosmological scales. Consequently, they serve as essential tools for studying the relationship between quantum mechanics, gravity, and the fundamental nature of spacetime itself.
		
\vspace{0.2cm}
\setlength{\parindent}{0pt}
		
Astrophysical evidence increasingly suggests that the centers of most galaxies harbor supermassive black holes (SMBHs). A striking example is the Virgo A galaxy M87, which hosts a BH that is approximately 1500 times more massive and 2000 times more distant than Sgr~A$^{\ast}$ at the center of the Milky Way \cite{a-5,a-6,a-7}. The complete characterization of BHs through parameters such as mass, angular momentum, and electric charge enables rigorous tests of GR predictions against observational data \cite{a-8,EHTL5,EHTL12,EHTL17,Do,GRAVITY,GRAVITY2020,Kocherlakota,Vagnozzi,a-10}. High-resolution imaging by the Event Horizon Telescope (EHT) has produced the first direct visualization of the M87 BH shadow, revealing distorted accretion disk structures and photon-ring features resulting from strong gravitational lensing \cite{a-8,a-9,a-10}. The distinction between stable and unstable light orbits around the photon sphere provides a precise diagnostic of the underlying spacetime geometry \cite{a-11,a-12}. Complementary evidence for BH existence and dynamics arises from gravitational-wave detections by LIGO/Virgo, which provide detailed information on horizon geometry, spin, and energy emission during BH mergers \cite{a-13}, thereby enabling a deeper understanding of compact object coalescence and strong-field gravitational dynamics. Together with horizon-scale electromagnetic observations by the Event Horizon Telescope, these results highlight the growing synergy between electromagnetic and gravitational-wave probes of black holes, ushering in an era of precision black hole astrophysics \cite{a-9,a-10,a-13}.
		
\vspace{0.2cm}
		
\setlength{\parindent}{0pt}
		
The study of BH-shadows has emerged as a particularly powerful observational tool for testing both GR and alternative gravity models \cite{a-14}. Shadows encode information not only about the spacetime geometry but also about accretion dynamics, photon trajectories \cite{a-15,a-16,a-17,a-18,a-19,a-20,a-21,a-22,a-23,newref1,newref2,newref3,newref4}, and the influence of additional matter distributions, such as dark matter (DM) halos \cite{,a-24,a-25,a-26,a-27,a-28,a-29,newref5,newref6,newref7,newref8}. DM, which is ubiquitous in galaxies and intergalactic medium, is inferred from a variety of astrophysical observations, including galaxy rotation curves, mass-to-light ratios, baryon acoustic oscillations, and the cosmic microwave background \cite{a-36,a-37,a-38,a-39,a-40}. While the effects of dark energy (DE) are generally subdominant near BHs, DM can significantly alter local spacetime geometry and particle dynamics, motivating detailed studies of DM spikes (DMSs) in the vicinity of supermassive black holes (SMBHs) \cite{a-24,a-41,a-42,a-43,a-44,a-45,a-46,a-47,a-48,a-50,a-51,a-52,newref9,newref10}. Gondolo and Silk \cite{a-56} initially modeled the formation of DM spikes via adiabatic BH growth in halos with initial density profiles $\rho \sim r^{-\gamma}$, predicting enhanced DM density near the BH horizon, $\rho \sim r^{-\gamma_{SP}}$. Subsequent works extended these models to incorporate general relativistic effects for Schwarzschild and Kerr BHs \cite{a-57,a-59}, revealing profound implications for gamma-ray signals and indirect detection of DM annihilation \cite{a-33,a-57,a-59}. These studies emphasize the critical role of DM distributions in shaping BH observational signatures and provide a potential link between particle physics and astrophysics. Moreover, DM spikes can influence the stability of photon orbits and the morphology of BH shadows, potentially yielding observable deviations from pure GR predictions that serve as probes of the DM profile near the event horizon \cite{a-15,a-17,a-25,a-26,a-33}.
		
\vspace{0.2cm}
		
\setlength{\parindent}{0pt}
		
Concurrently, Lorentz symmetry, a foundational principle of GR and particle physics, has been extensively examined in contexts where it may be broken or deformed. Frameworks including string theory \cite{b1}, loop quantum gravity \cite{b2}, Ho\v{r}ava-Lifshitz gravity \cite{b3}, and noncommutative field theory \cite{b4} suggest that Lorentz symmetry breaking (LSB) could manifest at high energy scales, potentially affecting BH physics. The Standard-Model Extension provides a robust phenomenological approach for studying LSB \cite{b5}, incorporating both GR and controlled violations of Lorentz invariance. In this context, the bumblebee model introduces a vector field $B_\mu$ that acquires a non-zero vacuum expectation value (VEV), spontaneously breaking Lorentz symmetry and deforming the spacetime geometry \cite{b1,b6,b7,b8,b9}. Bumblebee gravity has been extensively explored in BH solutions \cite{b10,hassan,b11,b12,b13,b14,b15,b16,b17,b18,b19,b20,b21,b22,b23,b24,b25,b26,b27,b28,b29,b30,b31,b32,b33,b34,b35,b36,b37}, wormholes \cite{b38}, cosmology \cite{b39,b40}, and gravitational waves \cite{b41,b42}. Further extensions involving the Kalb-Ramond (KR) field $B_{\mu\nu}$ \cite{b43,Duan,b44,b45,b46,b47,b48,b49,b50,b51,b52,b53,b54,b55,b56,b57,b58} provide alternative mechanisms for spontaneous LSB, influencing BH solutions, particle motion, shadow properties, and quasinormal mode spectra \cite{b59}. These developments underscore the importance of Lorentz-violating fields in exploring new physics near BHs and in strongly-curved spacetimes. The relationship between LSB and astrophysical observables, such as shadow shape distortions, precession of orbital planes, and modifications of accretion disk dynamics, opens new windows for testing fundamental symmetries in the strong-gravity regime \cite{b10,b12,b14,b21,b37,b42}.
		
\vspace{0.2cm}
		
\setlength{\parindent}{0pt}
		
The study of Anti-de Sitter (AdS) BHs connects these considerations to holographic duality and BH thermodynamics. Solutions of Einstein's equations with a negative cosmological constant underpin the AdS/CFT correspondence, linking gravitational dynamics in the bulk to conformal field theories on the boundary \cite{c1,c2,c3,c4}. The Schwarzschild-AdS BH exemplifies this class, exhibiting the Hawking-Page phase transition \cite{c5}, which can be interpreted as a confinement/deconfinement transition in the dual gauge theory \cite{c6}. Electrically charged and rotating extensions, such as Kerr-Newman-AdS BHs, introduce additional conserved quantities-angular momentum and electric charge-that affect horizon structure, ergoregion geometry, and thermodynamic properties \cite{c7,c8,c9,c10}. These BHs display van der Waals-Maxwell-like phase transitions \cite{c7,c9,c10}, with rotation and charge critically influencing specific heats, critical points, and stability criteria \cite{c11,c8,c13}. Detailed studies of these phase transitions provide insights into black hole microphysics, horizon thermodynamics, and the deep connections between gravitational dynamics and quantum field theories \cite{c3,c5,c6,c7,c9,c13}.
		
\vspace{0.2cm}
		
\setlength{\parindent}{0pt}
		
Integrating Lorentz-violating fields within Kerr-Newman-AdS geometries yields the so-called Kerr-Newman-AdS Bumblebee BHs \cite{c8,c14,c15,c16,c17,c18,c19,aa-6,c21,c22,c23,c24}. Here, the bumblebee vector field introduces preferred spacetime directions that deform the horizon and ergoregion, modify the Hawking temperature and Bekenstein-Hawking entropy, and impact the phase structure and energy extraction processes \cite{b9,b14,b21,b30,b36,c8,c14}. When coupled with DM spikes, these effects are further amplified near supermassive black holes (SMBHs), affecting shadow morphology, photon orbit stability, innermost stable circular orbits (ISCOs), accretion dynamics, and relativistic jet formation \cite{a-41,a-45,a-52,c14,c25,c26,c27,c28,c29,c30,c31,c32,c33,c34,c35,c36,c37,c38,c39,c40,c41,c42}.  
		
\vspace{0.2cm}
		
\setlength{\parindent}{0pt}
		
The study of ISCOs is particularly significant in understanding particle dynamics around Kerr-Newman-AdS Bumblebee BHs. The presence of DM spikes modifies the effective potential for equatorial motion, thereby altering the radius of the ISCO and impacting the efficiency of energy conversion in accretion processes \cite{a-44,a-46,a-50,a-52,c30,c32,c33}. Similarly, the Penrose process, which extracts rotational energy from the ergoregion, is influenced by both the LSB vector field and DM distribution, offering a theoretically rich framework for studying energy extraction in strongly curved spacetimes \cite{b14,b21,c8,c25,c29}. These mechanisms provide direct connections between BH rotation, spacetime anisotropies, and observationally relevant energy fluxes \cite{a-15,a-17,a-19,c26,c31,c34}.
		
\vspace{0.2cm}
		
\setlength{\parindent}{0pt}
		
Thin-disk accretion models provide another critical theoretical tool for examining energy release near BHs. Theoretical calculations of radial energy flux, temperature distribution, and luminosity spectra in the presence of Kerr-Newman-AdS Bumblebee geometries, coupled with DM spikes, enable the study of deviations from classical Novikov-Thorne predictions \cite{c30,c33,c36,c38}. Such studies reveal how Lorentz symmetry violation and surrounding matter distributions can significantly alter the radiative properties of the accretion disk, influencing theoretically predicted observational signatures \cite{b14,b20,b22,b35,c25,c31,c39}.
		
\vspace{0.2cm}
		
\setlength{\parindent}{0pt}
		
Hawking radiation sparsity is also an important theoretical aspect in this context. The combined effects of rotation, LSB, and DM density spikes modify the emission spectrum and particle flux near the horizon \cite{b12,b17,b25,b36,c19,aa-6,c28}. These modifications have profound implications for black hole thermodynamics, including the rate of mass loss, energy distribution of emitted quanta, and potentially observable signatures in high-energy astrophysical environments \cite{c3,c4,c5,c8,c14,c18}. By theoretically modeling these effects, one can explore fundamental aspects of semiclassical gravity in non-trivial spacetimes \cite{a-3,a-4,b6,b9,c1,c2,c5}.
		
\vspace{0.2cm}
		
\setlength{\parindent}{0pt}
		
In this work, we present a comprehensive investigation of the thermodynamic, optical, and dynamical properties of Kerr-Newman-AdS black holes within the Bumblebee gravity model, incorporating the effects of DM spikes, quintessence-like fields, and spontaneous Lorentz-symmetry breaking. We examine horizon and ergoregion deformation, BH shadow structure, sparsity of Hawking radiation, particle motion in the equatorial plane, ISCOs, thin-disk energy fluxes, and energy extraction processes. By systematically integrating the combined effects of rotation, charge, negative cosmological constant, DM distribution, and LSB, this study provides a unified framework for theoretical predictions and potential observational tests, advancing our understanding of the rich phenomenology of astrophysical BHs. The results presented here pave the way for precision observational and theoretical tests of GR, DM physics, and Lorentz symmetry in the strong-field regime, contributing to the ongoing effort to connect theory with observation in modern astrophysics.

This manuscript is structured as follows: In Section~\ref{sec:BKNBH}, we present the Bumblebee Kerr-Newman-AdS spacetime, analyzing the horizon structure and the effects of Lorentz violation, rotation, charge, and cosmological constant. Section~\ref{sec:Thermo} addresses thermodynamics, including Hawking temperature, entropy, heat capacity, and remnant formation, highlighting stability, phase transitions, and the role of Lorentz-symmetry breaking. Section~\ref{sec:sparsity} investigates Hawking radiation sparsity, while Section~\ref{sec:optical} studies photon trajectories, shadow morphology, and observable effects of black hole parameters. Section~\ref{subsec:ergosphere} examines the ergosphere’s shape and volume under Lorentz violation. In Section~\ref{sec:Penrose}, we analyze the Penrose energy extraction mechanism, and Section~\ref{subsec:PRT} quantifies the proper radial thickness of the ergosphere, showing its angular dependence and linear growth with spin via a near-horizon, slow-rotation expansion. Section~\ref{sec:conc} summarizes our results, emphasizing how the Lorentz-violating parameter modifies horizons, thermodynamics, radiation, photon dynamics, shadow features, ergoregion properties, and energy extraction, and outlines directions for future studies.

\section{The Bumblebee Kerr-Newman-AdS Black Holes}\label{sec:BKNBH}

In the framework of Lorentz-violating gravity, the so-called Bumblebee models introduce a dynamical vector field that spontaneously breaks Lorentz symmetry. When coupled to gravity, this mechanism modifies the standard spacetime geometry by introducing an anisotropic parameter that quantifies the degree of Lorentz symmetry breaking. Within this context, the corresponding static and spherically symmetric charged AdS black hole solution is described by the line element \cite{c34}
\begin{equation}\label{eq-2}
	ds^2 = -f(r)\,dt^2 + \frac{dr^2}{g(r)} + h(r)\left(d\theta^2 + \sin^2\theta\, d\phi^2\right),
\end{equation}
where the metric functions are given by
\begin{equation}
	f(r) = g(r) = \frac{1}{1-l} - \frac{2M}{r} + \frac{Q^2}{(1-l)^2 r^2} - \frac{\Lambda r^2}{3(1-l)},\qquad h(r) = r^2.
\end{equation}
Here, $M$ represents the black hole mass, $Q$ denotes its electric charge, $\Lambda$ is the cosmological constant associated with an asymptotically anti-de Sitter (AdS) spacetime, and $l$ is the Lorentz-violating parameter arising from the vacuum expectation value of the Bumblebee field. The function $h(r)$ maintains the conventional areal radius form. The presence of the parameter $l$ effectively deforms the standard Reissner-Nordström-AdS geometry by rescaling the gravitational coupling and the electromagnetic contribution, thereby encoding deviations from exact Lorentz invariance in the gravitational sector.

\begin{figure*}[htbp]
	\centering

	\begin{subfigure}{0.45\textwidth}
		\centering
		\includegraphics[width=\linewidth]{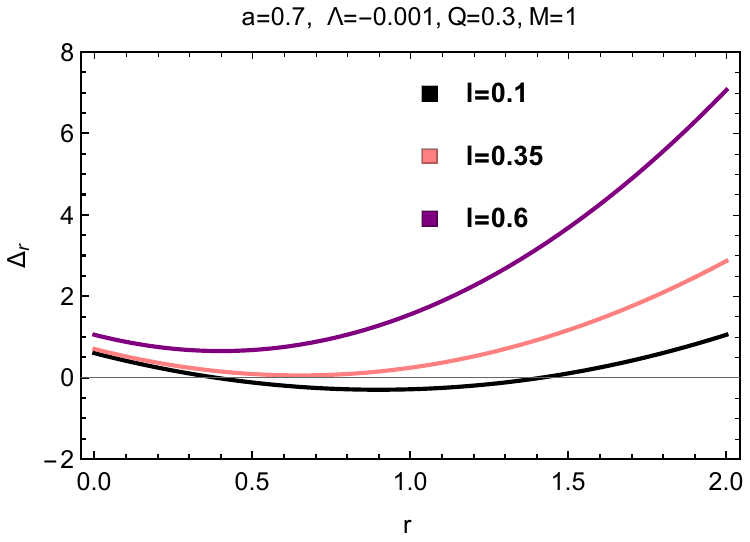}
		\caption{}
	\end{subfigure}
	\hfill
	\begin{subfigure}{0.45\textwidth}
		\centering
		\includegraphics[width=\linewidth]{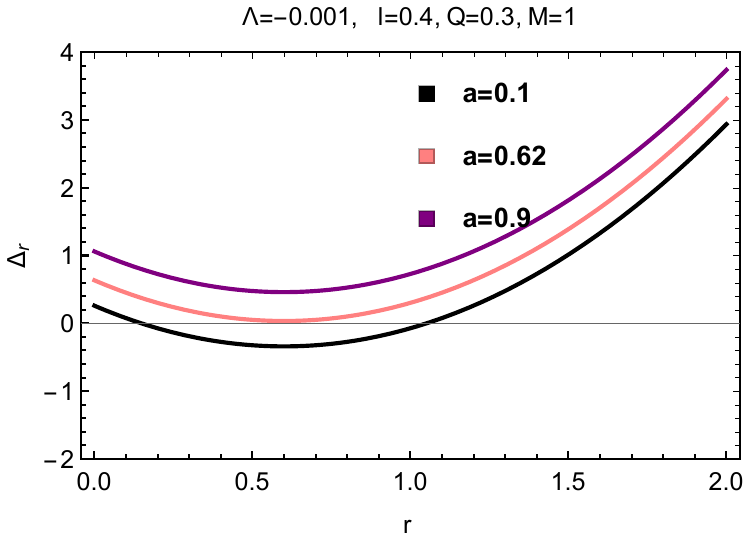}
		\caption{}
	\end{subfigure}
	
	\vspace{0.6em}
	
	\begin{subfigure}{0.45\textwidth}
		\centering
		\includegraphics[width=\linewidth]{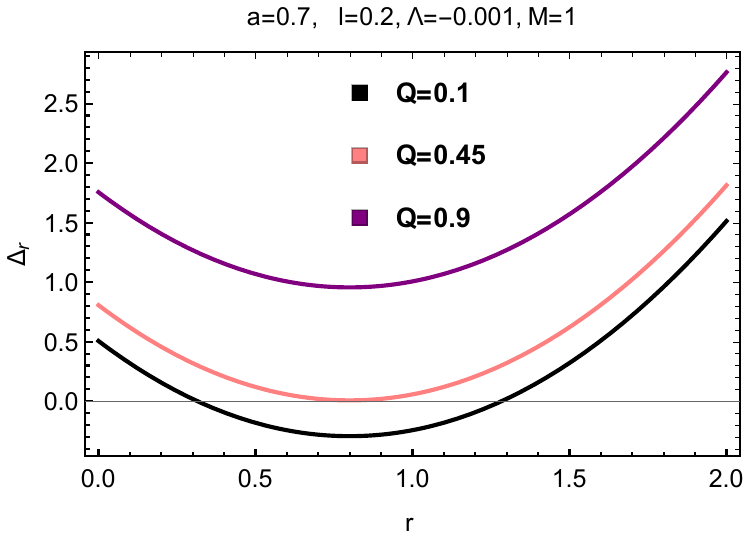}
		\caption{}
	\end{subfigure}
	
	\caption{Plot of $\Delta_r$ versus $r$ for different values of the black hole parameters, illustrating the effects of (a) Lorentz-violating parameter $l$, (b) rotation parameter $a$, and (c) electric charge $Q$.}
	\label{f-1}
\end{figure*}

To obtain the rotating counterpart of this geometry, one applies the standard Newman-Janis complex transformation to introduce rotation through the specific angular momentum $a$. The resulting line element, corresponding to the rotating Bumblebee Kerr-Newman-AdS black hole, takes the form
\begin{equation}\label{RotationMetric}
\begin{split}
ds^2 &= \frac{\Sigma}{\Delta_r}\,dr^2 + \frac{\Sigma}{\Delta_\theta}\,d\theta^2 + \frac{\Delta_\theta \sin^2\theta}{\Sigma} \left(\frac{a\,dt}{\Xi} - (r^2+a^2)\frac{d\phi}{\Xi}\right)^2\\
&- \frac{\Delta_r}{\Sigma}\left(\frac{dt}{\Xi} - a\sin^2\theta \frac{d\phi}{\Xi}\right)^2,
\end{split}
\end{equation}
where the metric functions are defined as
\begin{equation}
	\begin{split}
		\Delta_r &= r^2 - 2Mr + a^2 + \frac{Q^2}{(1-l)^2} - \frac{\Lambda r^2}{3(1-l)}(r^2 + a^2) + \frac{l}{1-l} r^2,\\
		\Delta_\theta &= 1 + \frac{\Lambda}{3(1-l)} a^2 \cos^2\theta,\quad \Xi = 1 + \frac{\Lambda}{3(1-l)} a^2,\quad \Sigma = r^2 + a^2 \cos^2\theta.
	\end{split}
\end{equation}
This solution reduces to several well-known limits under specific parameter choices. When $l=0$, the metric reproduces the standard Kerr-Newman-AdS black hole. Setting $l=\Lambda=0$ yields the asymptotically flat Kerr-Newman geometry, and further imposing $Q=\Lambda=l=0$ reduces the metric to the classical Kerr black hole. Hence, the parameter $l$ quantifies the departure from local Lorentz symmetry and provides a direct means to investigate the influence of Lorentz-violating effects on the AdS curvature. The function $\Delta_r$ plays a central role in determining the horizon structure of the black hole. 
\begin{figure*}[htbp]
	\centering
	
	\begin{subfigure}{0.45\textwidth}
		\centering
		\includegraphics[width=\linewidth]{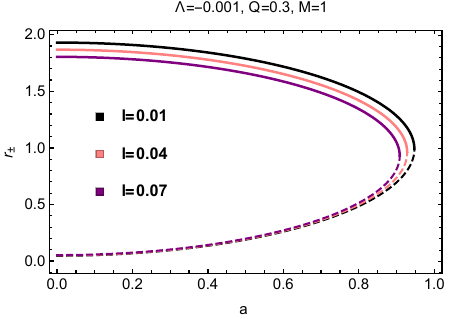}
		\caption{}
	\end{subfigure}
	\hfill
	\begin{subfigure}{0.45\textwidth}
		\centering
		\includegraphics[width=\linewidth]{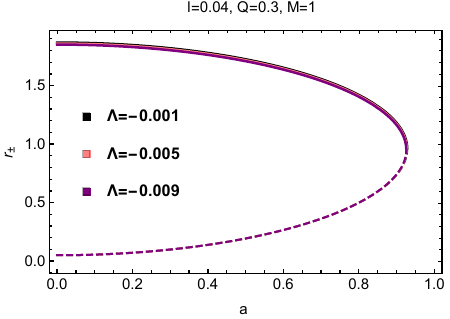}
		\caption{}
	\end{subfigure}
	
	\vspace{0.6em}
	
	\begin{subfigure}{0.45\textwidth}
		\centering
		\includegraphics[width=\linewidth]{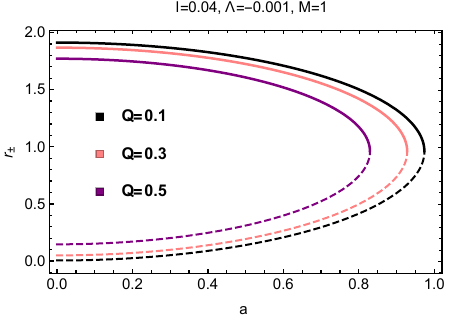}
		\caption{}
	\end{subfigure}
	
	\caption{Event horizon ($r_+$, thick line) and Cauchy horizon ($r_-$, dashed line) as functions of the rotation parameter $a$ for different values of (a) $l$, (b) $\Lambda$, and (c) $Q$. The plots show how Lorentz-violating effects, cosmological curvature, and charge modify the horizon structure.}
	\label{f-rpm}
\end{figure*}

For the rotating Bumblebee Kerr-Newman-AdS spacetime, the condition $\Delta_r(r_h) = 0$ determining the horizon radii can be written in the general form of a depressed quartic equation,
\begin{equation}\label{e-2-5}
\alpha\, r^4 + b\, r^2 - 2\, M\, r + c = 0,
\end{equation}
where the coefficients are
\begin{equation}\label{e-2-6}
	\alpha = -\frac{\Lambda}{3(1-l)}, \qquad b = \frac{1 - \frac{\Lambda a^2}{3}}{1-l}, \qquad c = a^2 + \frac{Q^2}{(1-l)^2}.
\end{equation}
The quartic structure of this equation implies a rich variety of possible horizon configurations, depending on the relative magnitudes of $(M, Q, a, \Lambda, l)$. The presence of the Lorentz-violating parameter $l$ modifies both the curvature and charge contributions, thereby shifting the conditions under which extremal or non-extremal configurations arise.

An extremal black hole corresponds to the case in which $\Delta_r$ possesses a double real root $r_{\text{ext}}$, at which the inner and outer horizons coincide. Imposing the simultaneous conditions $\Delta_r(r_{\text{ext}}) = 0$ and $\Delta_r'(r_{\text{ext}}) = 0$ and eliminating $M$ yields the constraint
\begin{equation}
3\,\alpha\, r_{\text{ext}}^4 + b\, r_{\text{ext}}^2 - c = 0,
\end{equation}
which gives
\begin{equation}
r_{\text{ext}}^2 = \frac{-b + \sqrt{b^2 + 12\, \alpha\, c}}{6\, \alpha}, \qquad M_{\text{ext}} = 2\,\alpha\, r_{\text{ext}}^3 + b\, r_{\text{ext}}.
\end{equation}
For an AdS background with $\Lambda < 0$, we have $\alpha > 0$, ensuring $r_{\text{ext}}^2 > 0$ and thereby preserving the physicality of the extremal radius. In this case, $\Delta_r(r)$, being a quartic polynomial that opens upwards with $\Delta_r(0) = c > 0$ and $\Delta_r(r) \to +\infty$ as $r \to \infty$, can exhibit three distinct horizon configurations: (i) two distinct real positive roots corresponding to a non-extremal black hole with separate event and Cauchy horizons; (ii) a single double positive root for the extremal configuration; or (iii) no real positive roots, corresponding to a naked singularity.

Therefore, for $M > M_{\text{ext}}$, the spacetime possesses two distinct horizons, with $0 < r_- < r_+$. For $M = M_{\text{ext}}$, the horizons coalesce at $r = r_{\text{ext}}$, marking the transition to extremality. Finally, when $M < M_{\text{ext}}$, the absence of real positive roots signals the breakdown of cosmic censorship, exposing the central singularity to external observers. The dependence of these regimes on the Lorentz-violating parameter $l$ is particularly noteworthy: increasing $l$ effectively alters the curvature scale and charge contribution, shifting the extremality condition and influencing the stability and causal structure of the black hole in a physically measurable manner.

Figure~\ref{f-1} illustrates the behavior of $\Delta_r$ as a function of the radial coordinate $r$ for various parameter choices. Panel (a) shows the effect of the Lorentz-violating parameter $l$, panel (b) exhibits the influence of the rotation parameter $a$, and panel (c) highlights the dependence on the electric charge $Q$. In each case, the zeros of $\Delta_r$ correspond to the radial positions of the horizons. Setting $\Delta_r = 0$ yields a quartic equation in $r$, whose roots determine the causal structure of the spacetime. Typically, two of these roots are complex and hence unphysical, while the remaining two are real and positive. The smaller of these two real roots corresponds to the inner (Cauchy) horizon $r_-$, while the larger one defines the outer (event) horizon $r_+$. The event horizon represents the one-way boundary beyond which no signal or particle can escape, while the Cauchy horizon separates regions of deterministic and non-deterministic evolution inside the black hole. The relative positions of these horizons govern the thermodynamic and causal properties of the system. 

Figure~\ref{f-rpm} depicts the variation of both the event horizon ($r_+$, thick line) and the Cauchy horizon ($r_-$, dashed line) with respect to different parameters. In panel (a), the horizons are shown as functions of the rotation parameter $a$ for several values of the Lorentz-violating parameter $l$; in panel (b), the effect of varying the cosmological constant $\Lambda$ is presented; and in panel (c), the dependence on the electric charge $Q$ is displayed. As expected, increasing the rotation parameter $a$ or the charge $Q$ tends to reduce the separation between the two horizons, while Lorentz-violating effects shift their positions through an effective rescaling of the gravitational potential.

\section{Thermodynamic Properties of Bumblebee Kerr-Newman-AdS Black Holes}\label{sec:Thermo}

Hawking radiation originates as a semiclassical quantum phenomenon arising in the curved spacetime near a black hole's event horizon, where vacuum fluctuations produce virtual particle-antiparticle pairs. The particle separation mechanism, facilitated by the strong gravitational gradient, enables one particle to escape as radiation while its counterpart falls into the black hole, reducing the mass. The radiation intensity is inversely proportional to the black hole mass; thus, low-mass black holes radiate with exceptional intensity. This effect is particularly pronounced in micro black holes, where the extreme curvature at the horizon dramatically enhances pair production, leading to an exponentially higher temperature and accelerated mass loss \cite{c18,evap2,c21}.  

The Hawking temperature for the Bumblebee Kerr-Newman-AdS black hole, incorporating rotation \(a\), charge \(Q\), cosmological constant \(\Lambda\), and Lorentz-violating parameter \(l\), is given by
\begin{equation}
\begin{aligned}
&T_H = \frac{\Delta'_r(r)}{4\pi (a^2+r^2)}\Big|_{r=r_h}
    = -\frac{r_h \left(a^2 \Lambda + 2\Lambda r_h^2 - 3\right)}
           {6\pi (1-l)\left(a^2 + r_h^2\right)} \\
           \\
& - \frac{3 a^2 l^2 + a^2 \Lambda l r_h^2 - 6 a^2 l - a^2 \Lambda r_h^2
        + 3 a^2 + \Lambda l r_h^4 - 3 l r_h^2 + 3 Q^2
        - \Lambda r_h^4 + 3 r_h^2}
        {12\pi (1-l)^2 \left(a^2 + r_h^2\right)\, r}
\end{aligned}
\label{eq:TH}
\end{equation}
\begin{figure*}[htbp]
	\centering
	
	\begin{subfigure}{0.45\textwidth}
		\centering
		\includegraphics[width=\linewidth]{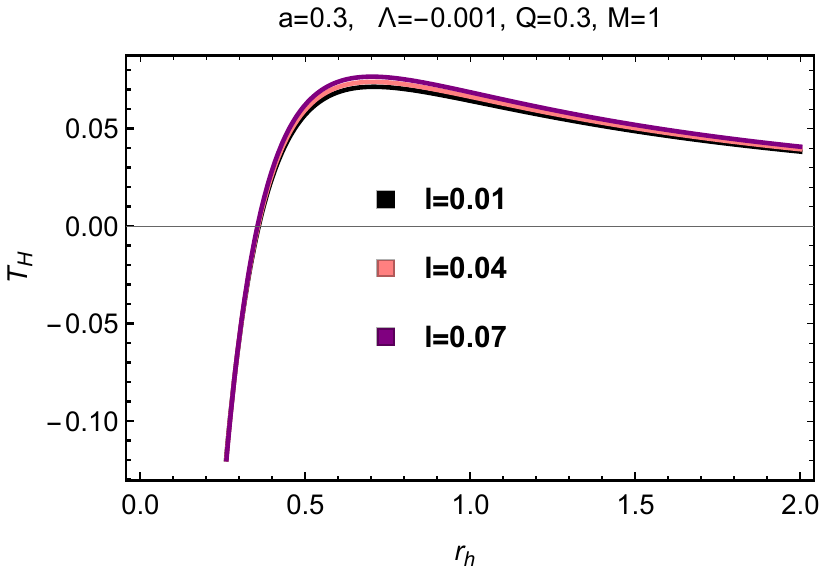}
		\caption{}
	\end{subfigure}
	\hfill
	\begin{subfigure}{0.45\textwidth}
		\centering
		\includegraphics[width=\linewidth]{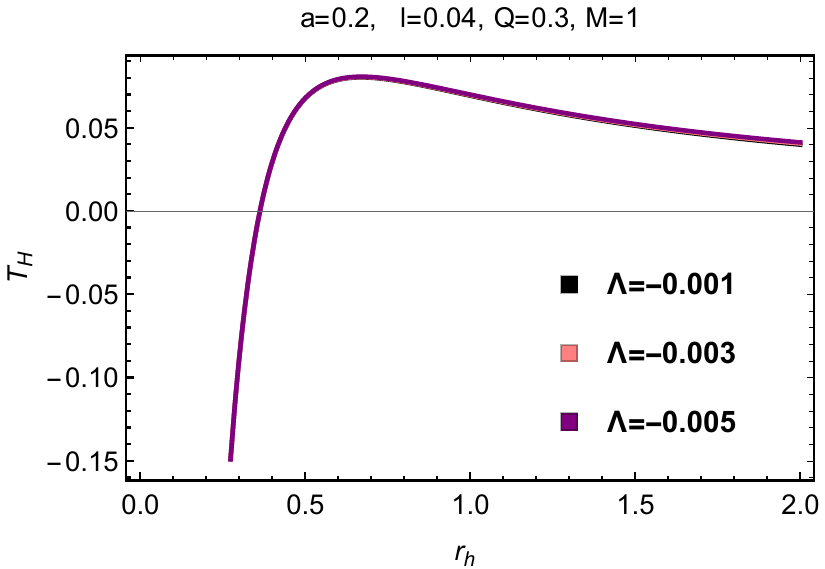}
		\caption{}
	\end{subfigure}
	
	\vspace{0.6em}
	
	\begin{subfigure}{0.45\textwidth}
		\centering
		\includegraphics[width=\linewidth]{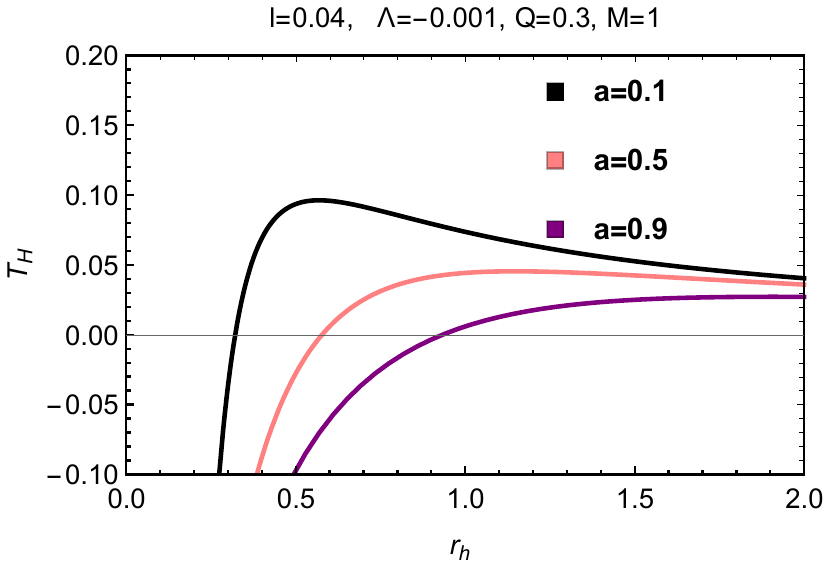}
		\caption{}
	\end{subfigure}
	\hfill
	\begin{subfigure}{0.45\textwidth}
		\centering
		\includegraphics[width=\linewidth]{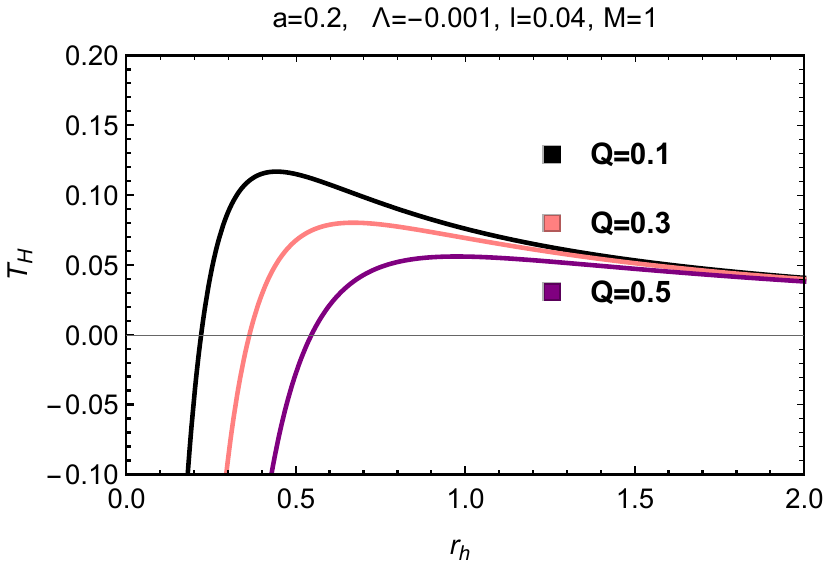}
		\caption{}
	\end{subfigure}
	
	\caption{Hawking temperature \(T_H\) versus horizon radius \(r_h\) for different black hole parameters: (a) varying \(l\), (b) varying \(\Lambda\), (c) varying \(a\), (d) varying \(Q\). Extremal points shift with rotation, charge, and Lorentz-violating parameters, defining critical radii for evaporation and remnant formation.}
	\label{f-TH}
\end{figure*}

Figure~\ref{f-TH} displays the temperature profile as a function of \(r_h\) for different values of the parameters \(l\), \(\Lambda\), \(a\), and \(Q\). Small-radius black holes reach extremely high temperatures, while large-radius black holes approach lower, asymptotic temperatures, reflecting the inverse relation between temperature and horizon radius. Panels (c) and (d) reveal that increasing rotation and charge not only elevates the extremal horizon radius but also shifts the onset of remnant formation, demonstrating their stabilizing effect on micro black holes. The location of extrema and the slope of the temperature curves provide a clear indication of critical points corresponding to phase transitions.

The remnant configuration, characterized by vanishing temperature \(T_H=0\), defines an extremal black hole with finite mass and radius. The remnant radius is determined analytically as
\begin{equation}
\begin{split}
 r_{\text{remnant}}&=\frac{1}{\sqrt{6\Lambda}}
\sqrt{	- a^2 \Lambda	+	\sqrt{\tilde{\mu}}	+ 3}\qquad \text{where}\\
\\
&\tilde{\mu}=a^4\Lambda^2+ 6\Lambda\!\left[a^2(6l-7) -\frac{6Q^2}{1-l}\right]+ 9.
\end{split}
\end{equation}
Substituting \(r_{\text{remnant}}\) into the mass function derived from \(\Delta_r=0\) yields the remnant mass:
\begin{equation}
\begin{aligned}
M_{remnant} &= 
\frac{ \left( 2 a^2 (-\Lambda)  - \sqrt{\tilde{\mu}}+ 6 \right) }
     { 9 \sqrt{6} (1-l)}
\sqrt{
\frac{
    a^2 \Lambda - \sqrt{\tilde{\mu}}- 3
}{
    486 \Lambda (1-l)^3
}
}.
\end{aligned}
\label{eq:Mrem}
\end{equation}
Table~\ref{tab-1} quantifies the remnant radius and mass for various rotation parameters. The growth of both quantities with \(a\) confirms that rotation enhances black hole stability, postpones complete evaporation, and defines extremal configurations that serve as thermodynamically stable endpoints of Hawking radiation.

\begin{figure*}[htbp]
	\centering
	
	\begin{subfigure}{0.45\textwidth}
		\centering
		\includegraphics[width=\linewidth]{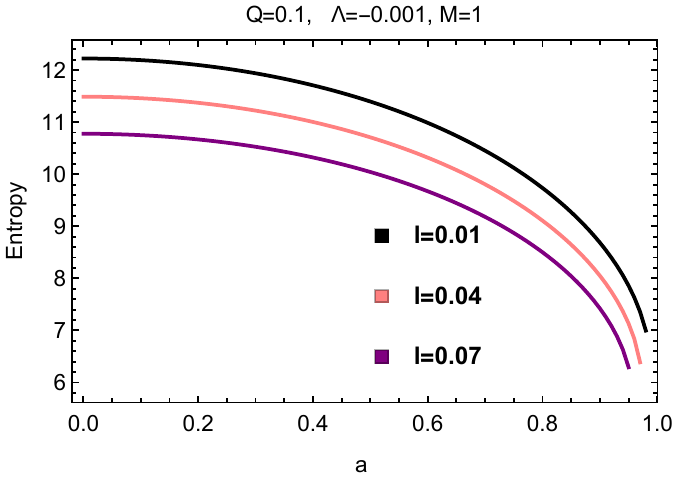}
		\caption{}
	\end{subfigure}
	\hfill
	\begin{subfigure}{0.45\textwidth}
		\centering
		\includegraphics[width=\linewidth]{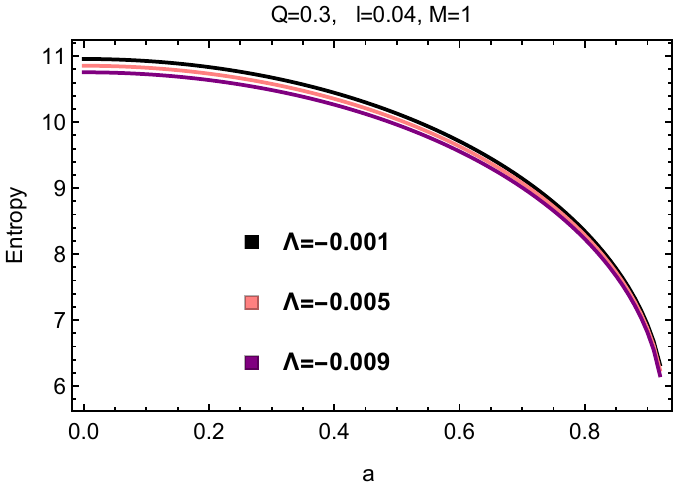}
		\caption{}
	\end{subfigure}
	
	\vspace{0.6em}
	
	\begin{subfigure}{0.45\textwidth}
		\centering
		\includegraphics[width=\linewidth]{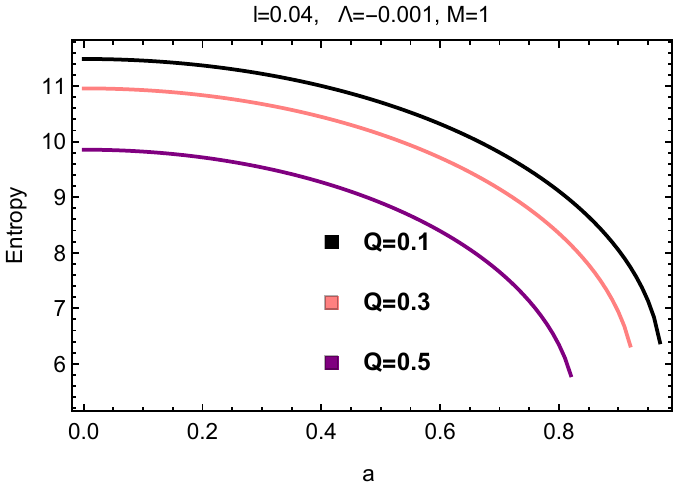}
		\caption{}
	\end{subfigure}
	
	\caption{Entropy \(S\) versus rotation parameter \(a\) for varying parameters: (a) \(l\), (b) \(\Lambda\), (c) \(Q\). The decreasing trend with \(a\) correlates with shrinking horizon area, while parameter variations shift the entropy curves, revealing effects of Lorentz violation, cosmological constant, and the charge.}
	\label{f-entropy}
\end{figure*}

\begin{table}[H]
\centering
\captionsetup{font=small, labelfont=bf}
\caption{ Remnant mass and remnant radius for different rotation parameters \(a\).}
\rowcolors{2}{lightgray}{white}
\begin{tabular}{ccc}
\toprule
\textbf{Rotation Parameter \(a\)} & \textbf{Remnant Radius \(r_{\text{remnant}}\)} & \textbf{Remnant Mass \(M_{\text{remnant}}\)} \\
\midrule
0.0 & 0.105351 & 0.117143 \\
0.1 & 0.141648 & 0.157650 \\
0.2 & 0.216401 & 0.241517 \\
0.3 & 0.301678 & 0.338237 \\
0.4 & 0.389859 & 0.439877 \\
0.5 & 0.478498 & 0.544211 \\
0.6 & 0.566532 & 0.650503 \\
0.7 & 0.653386 & 0.758504 \\
0.8 & 0.738710 & 0.868158 \\
0.9 & 0.822277 & 0.979492 \\
\bottomrule
\end{tabular}
\label{tab-1}
\end{table}
The event horizon area, corresponding to the boundary beyond which causal signals cannot escape, is
\begin{equation}
	A=\int_0^\pi\int_0^{2\pi}\sqrt{g_{\theta\theta}g_{\phi\phi}}\,d\theta d\phi\Big|_{r=r_+}=12 \pi  \Big(\frac{(1-l) \left(a^2+r_+^2\right)}{a^2 \Lambda -3 l+3}\Big),
\end{equation}
reducing to the Schwarzschild expression \(A=4\pi r^2\) for \(l=\Lambda=a=0\). The corresponding entropy is
\begin{equation}
	S=\frac{A}{4}=3 \pi   \Big(\frac{(1-l) \left(a^2+r_+^2\right)}{a^2 \Lambda -3 l+3}\Big),
\end{equation}
depicted as a function of \(a\) in Fig.~\ref{f-entropy}. Entropy decreases with increasing rotation due to the reduction in horizon area for fixed mass, whereas variations in \(l\), \(\Lambda\), and \(Q\) shift the curves, reflecting the contributions of Lorentz violation, electric charge, and the cosmological constant to the microphysical degrees of freedom of the horizon.

\begin{figure*}[htbp]
	\centering
	
	\begin{subfigure}{0.45\textwidth}
		\centering
		\includegraphics[width=\linewidth]{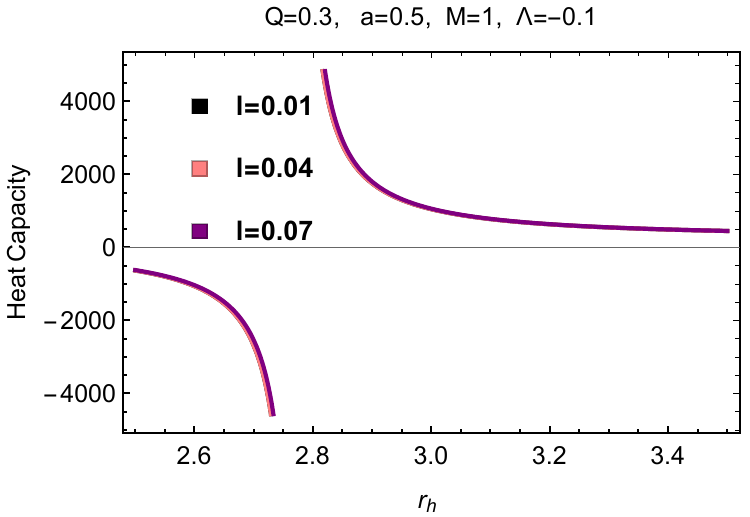}
		\caption{}
	\end{subfigure}
	\hfill
	\begin{subfigure}{0.45\textwidth}
		\centering
		\includegraphics[width=\linewidth]{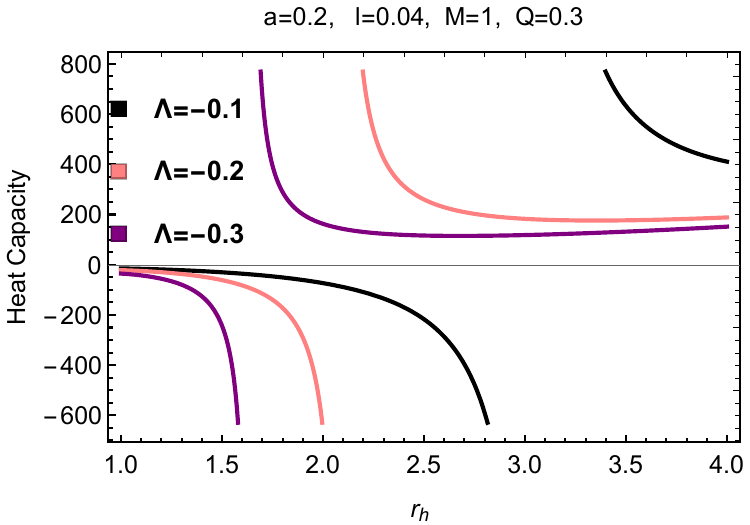}
		\caption{}
	\end{subfigure}
	
	\vspace{0.6em}
	
	\begin{subfigure}{0.45\textwidth}
		\centering
		\includegraphics[width=\linewidth]{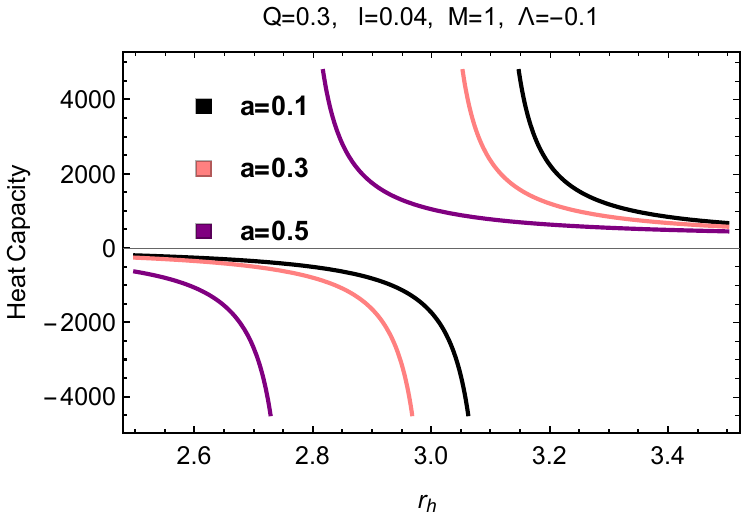}
		\caption{}
	\end{subfigure}
	\hfill
	\begin{subfigure}{0.45\textwidth}
		\centering
		\includegraphics[width=\linewidth]{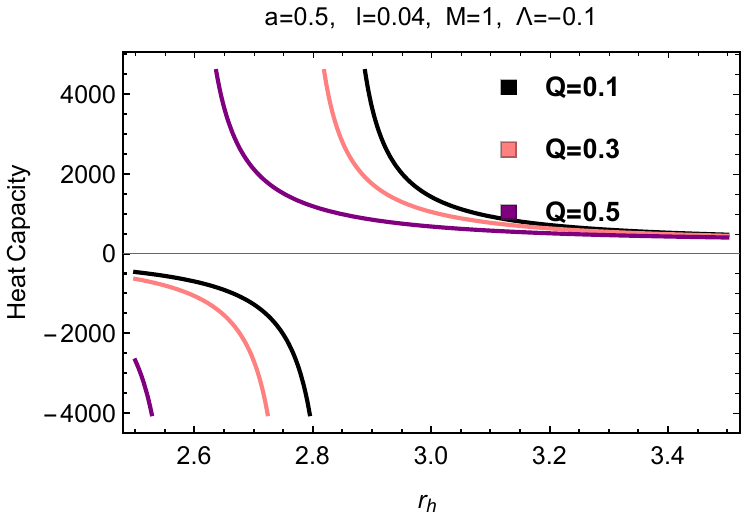}
		\caption{}
	\end{subfigure}
	
	\caption{Heat capacity \(C_V\) versus horizon radius \(r_h\) for varying parameters: (a) \(l\), (b) \(\Lambda\), (c) \(a\), (d) \(Q\). Positive values correspond to stable phases, negative values to instability, with divergences defining critical radii of phase transitions influenced by rotation, charge, and Lorentz-violating effects.}
	\label{f-CV}
\end{figure*}

The heat capacity of the black hole \cite{c18,c19,c21,c22,c23,c24}, a measure of its response to temperature changes, is defined by \cite{c24}
\begin{equation}
C_V=\frac{\partial M}{\partial T_H}=\frac{\frac{\partial M}{\partial r}}{\frac{\partial T_H}{\partial r}}.
\end{equation}
For the Bumblebee Kerr-Newman-AdS black hole, it is explicitly
\begin{equation}
\begin{aligned}
C_V = {} &
\frac{2\,\pi\, (a^2+r^2)^2\, \tilde{p}}
{%
a^4  \big( 3l + \Lambda r^2 - 3 \big)
+ a^2 \Big(  r^2 (9l + 8\Lambda r^2 - 12) -  \frac{3Q^2}{1-l}\Big)+\tilde{q}
},
\\[6pt]
&\tilde{p}=\Lambda r^2 (a^2 + 3 r^2) +3 \big(a^2 (1-l)- r^2 + \frac{Q^2}{(1-l) } \big),
\\[6pt]
&\tilde{q}=3 r^4 (\Lambda r^2 + 1)- \frac{9 Q^2 r^2}{1-l}.
\label{CV}
\end{aligned}
\end{equation}
Figure~\ref{f-CV} shows the heat capacity as a function of \(r_h\). Regions with positive \(C_V\) correspond to thermodynamically stable configurations, while negative \(C_V\) indicate instability and runaway evaporation. Divergences of \(C_V\) define critical radii separating stable and unstable phases. Lorentz violation from the Bumblebee field modifies these divergences, softening the gravitational curvature near the horizon and slightly shifting critical points to larger radii, thereby increasing the domain of stability and affecting extremal black hole configurations. Analytically, the critical horizon radius \(r_c\) corresponding to phase transition points is determined by solving
\begin{equation}
\frac{\partial T_H}{\partial r_h}=0,
\end{equation}
which yields
\begin{equation}
r_c = \sqrt{\frac{3 \left(-a^2 (1-l) - \frac{Q^2 }{1-l}+  r_h^2\right)}{\Lambda  (a^2 + 3 r_h^2)}}.
\end{equation}
The corresponding extremal mass \(M_{\rm extremal}\) at this critical point is obtained from \(\Delta_r(r_c)=0\), giving
\begin{equation}
M_{\rm extremal} = \frac{1}{2r_c} \left[r_c^2 + a^2 + Q^2 + \frac{\Lambda}{3} r_c^2 (r_c^2 + a^2)\right].
\end{equation}
These analytic expressions allow precise characterization of thermodynamic critical points, define boundaries between stable and unstable phases, and quantify how rotation, charge, Lorentz violation, and cosmological constant influence extremal configurations. In the small-radius limit, scaling relations are \(T_H \sim 1/r_h\), \(S \sim r_h^2\), and \(C_V \sim -r_h^2\), indicating unstable micro black holes undergoing rapid evaporation. For large horizons, \(T_H \sim r_h\), \(S \sim r_h^2\), and \(C_V \sim r_h^2\), implying thermodynamic stability. The Bumblebee field shifts these relations subtly, enlarging the domain of stability and redefining extremal radii. Thus, the Bumblebee Kerr-Newman-AdS black hole exhibits a rich thermodynamic structure where rotation, charge, cosmological constant, and Lorentz violation determine horizon geometry, extremal configurations, critical points, remnant properties, and phase transitions. These analytic insights provide a fully quantitative framework for understanding black hole evolution and stability in the presence of Lorentz symmetry breaking.

\section{Sparsity of Hawking Radiation}\label{sec:sparsity}

In this section, the sparse nature of Hawking radiation is analyzed for Bumblebee Kerr-Newman-AdS black holes. Although black holes are often treated as ideal blackbody emitters, their radiation occurs in discrete quanta, with each emission separated by a finite time interval. The temporal separation between emissions is typically much longer than the characteristic oscillation period of the emitted radiation, which is a direct consequence of the quantum nature of the radiation. The sparsity parameter provides a quantitative measure of the temporal gap between successive emissions \cite{c40,c41,c42}. The sparsity parameter $\tilde{\eta}$ is defined as
\begin{equation}
	\tilde{\eta}=\frac{C}{g}\Big(\frac{\lambda_t^2}{A_{eff}}\Big),
\end{equation}
where $C$ is a dimensionless numerical constant, $g$ is the degeneracy factor associated with the spin of the emitted particle, $\lambda_t = 2\pi/T_H$ is the thermal wavelength corresponding to the Hawking temperature, and $A_{eff}$ is the effective area of the black hole relevant for the emission process. This definition compares the characteristic wavelength of the emitted quanta to the geometric scale of the horizon, thereby providing a measure of how discrete the radiation is relative to an ideal blackbody emission.

\begin{figure*}[htbp]
	\centering
	
	\begin{subfigure}{0.45\textwidth}
		\centering
		\includegraphics[width=\linewidth]{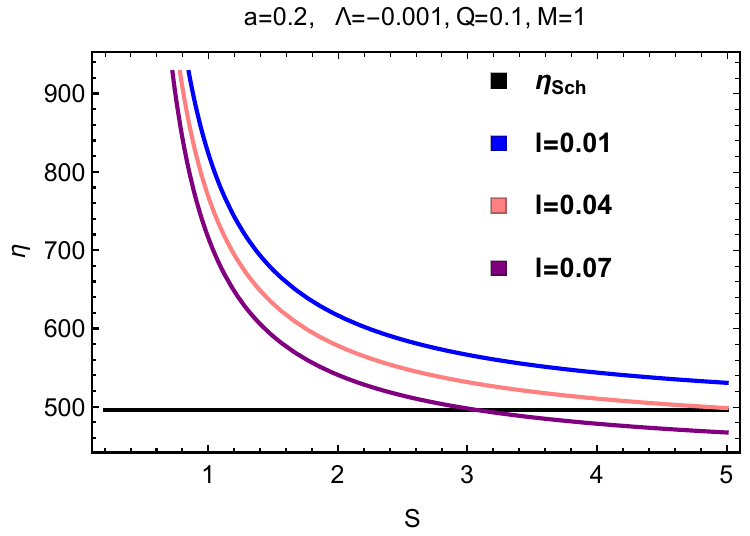}
		\caption{}
	\end{subfigure}
	\hfill
	\begin{subfigure}{0.45\textwidth}
		\centering
		\includegraphics[width=\linewidth]{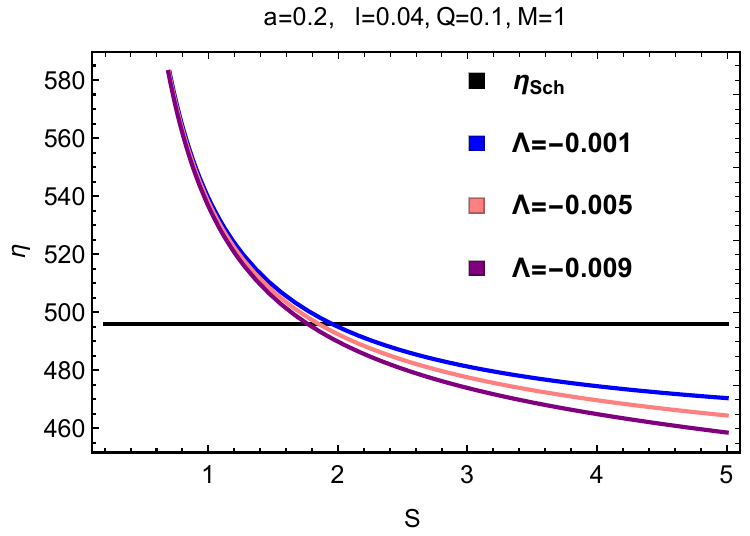}
		\caption{}
	\end{subfigure}
	
	\vspace{0.6em}
	
	\begin{subfigure}{0.45\textwidth}
		\centering
		\includegraphics[width=\linewidth]{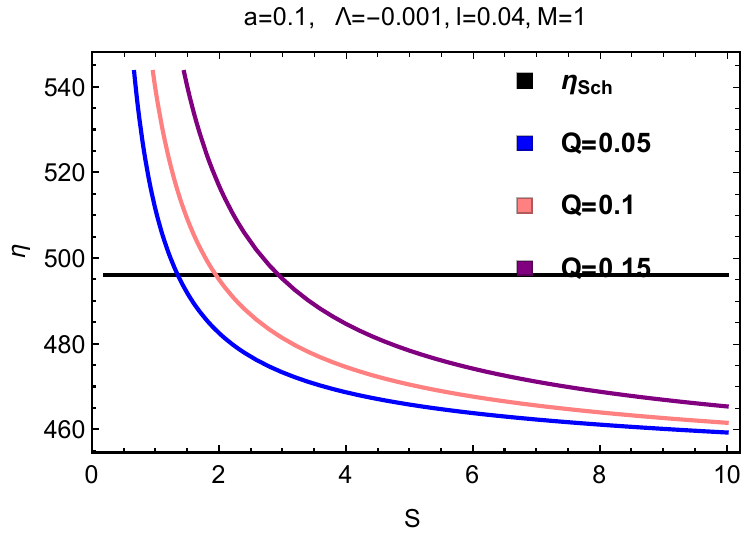}
		\caption{}
	\end{subfigure}
	
	\caption{Plot of the sparsity parameter $\eta$ versus entropy $S$ for different values of the black hole parameters, highlighting the dependence on (a) Lorentz violation $l$, (b) cosmological constant $\Lambda$, and (c) electric charge $Q$.}
	\label{f-sparsity}
\end{figure*}

Figure~\ref{f-sparsity} shows the variation of $\tilde{\eta}$ as a function of the black hole entropy $S$, which represents the horizon area and indirectly the black hole mass. The parameter $\tilde{\eta}$ decreases with increasing entropy, indicating that larger black holes emit radiation more continuously, while smaller black holes with lower entropy produce more discrete and intermittent quanta. Small black holes have high surface gravity and higher Hawking temperatures, leading to emission in highly irregular bursts. Larger black holes possess lower temperatures, which results in a more uniform, quasi-continuous radiation pattern.

The presence of the Bumblebee Lorentz-violating field modifies these characteristics. For a fixed entropy, $\tilde{\eta}$ remains larger than in the conventional Kerr-Newman-AdS black hole. The Bumblebee field introduces a preferred direction in spacetime, which reduces the effective surface gravity at the horizon and lowers the Hawking temperature $T_H$. Consequently, the thermal wavelength $\lambda_t$ increases and the emission rate decreases, resulting in a more discrete radiation spectrum. This effect emphasizes the influence of Lorentz symmetry breaking on the microscopic properties of Hawking radiation.

Rotation $a$, electric charge $Q$, and the cosmological constant $\Lambda$ also contribute to the structure of $\tilde{\eta}$. Rotation alters the near-horizon geometry through frame-dragging effects, modifying the effective emission area. Electric charge generates a repulsive effect that changes the surface gravity, and the cosmological constant introduces an asymptotic curvature that affects the global spacetime structure. These factors, combined with the Lorentz-violating parameter $l$, determine the hierarchy of sparsity across black hole configurations, from highly discrete emissions in micro black holes to near-continuous emissions in massive, high-entropy black holes.

Analytic scaling relations provide further insight. In the small-radius limit $r_h \ll 1$, $\lambda_t^2 \sim r_h^{-2}$ and $A \sim r_h^2$, yielding $\tilde{\eta} \sim r_h^{-4}$. This indicates that micro black holes experience highly discrete emission separated by long intervals. In the large-radius limit $r_h \gg 1$, $\lambda_t^2 \sim r_h^2$ and $A \sim r_h^2$, giving $\tilde{\eta} \sim \text{constant}$ or weakly varying with $r_h$, consistent with continuous emission. The Bumblebee field shifts these scaling relations, extending the small-radius regime of high sparsity and modifying the asymptotic behavior for large horizons, effectively redefining extremal radii and emission thresholds.

Physically, the enhanced sparsity has implications for the evolution of black holes, particularly primordial or microscopic ones. Emission in discrete, intermittent quanta affects the energy loss rate, potentially producing observable bursts of high-energy particles. Lorentz symmetry violation enhances the discreteness, providing a potential signal of beyond-standard-model physics in the quantum structure of spacetime.

Figure~\ref{f-sparsity} confirms that Bumblebee Kerr-Newman-AdS black holes occupy a higher sparsity regime than their standard counterparts. The combined influence of charge, rotation, cosmological constant, and Lorentz-violating background reshapes the temporal emission pattern, offering a quantitative framework to understand the discrete and nonclassical characteristics of Hawking radiation.

\section{Optical Properties of Bumblebee Kerr-Newman-AdS Black Holes}\label{sec:optical}

Our investigation now focuses on analyzing the propagation of photons in the vicinity of a Bumblebee Kerr-Newman-AdS black hole, as described by the stationary and axisymmetric metric given in Eq.~\eqref{RotationMetric}. This metric depends explicitly on the radial coordinate $r$ and the polar angle $\theta$, with rotation about the axis $\theta=0$, and includes contributions from the mass $M$, spin parameter $a$, electric charge $Q$, cosmological constant $\Lambda$, and Lorentz-violating parameter $l$. The functions $\Delta_r$, $\Delta_\theta$, $\Sigma$, and $\Xi$ encode the intricate interplay of these physical quantities and set the stage for the rich geodesic structure we explore. The geodesic motion of test particles, including photons, is conveniently described via the Lagrangian
\begin{equation}
2\mathcal{L} = g_{\mu\nu} \dot{x}^{\mu} \dot{x}^{\nu},
\end{equation}
where the overdot denotes differentiation with respect to the affine parameter $\lambda$. For timelike, null, and spacelike trajectories, $2\mathcal{L}$ takes the values $-1$, $0$, and $1$, respectively. In the photon case, we set $2\mathcal{L}=0$, reflecting their null geodesics, which represent the paths along which light propagates. To obtain the equations of motion, we employ the Hamilton-Jacobi formalism, whereby the canonical Hamiltonian $\mathcal{H}$ and Jacobi action $S$ satisfy
\begin{equation}\label{HamiltonJacobiEq}
\mathcal{H} = -\frac{\partial S}{\partial \lambda} = \frac{1}{2} g_{\mu\nu} \frac{\partial S}{\partial x^\mu} \frac{\partial S}{\partial x^\nu} = 0.
\end{equation}
Because the metric is independent of the coordinates $t$ and $\phi$, these directions correspond to Killing vectors $\xi^\mu_{(t)} = \delta^\mu_t$ and $\xi^\mu_{(\phi)} = \delta^\mu_\phi$, which generate conserved quantities associated with the system's symmetries. Specifically, these constants of motion are the total energy $E$ and the axial angular momentum $L_z$, which satisfy
\begin{equation}
E = -\frac{\partial S}{\partial t} = - g_{tt}\dot{t} - g_{t\phi} \dot{\phi}, \qquad 
L_z = \frac{\partial S}{\partial \phi} = g_{t\phi} \dot{t} + g_{\phi\phi} \dot{\phi}.
\end{equation}
Solving these relations for $\dot{t}$ and $\dot{\phi}$ yields
\begin{equation}
\dot{t} = \frac{g_{\phi\phi} E + g_{t\phi} L_z}{g_{t\phi}^2 - g_{tt} g_{\phi\phi}}, \qquad
\dot{\phi} = - \frac{g_{t\phi} E + g_{tt} L_z}{g_{t\phi}^2 - g_{tt} g_{\phi\phi}}.
\end{equation}
The separability of the Hamilton-Jacobi equation in this spacetime is achieved by expressing the Jacobi action as
\begin{equation}\label{JacobiAction}
S = - E t + L_z \phi + \int_0^r \frac{\sqrt{R(r')}}{\Delta_r} dr' + \int_0^\theta \sqrt{\Theta(\theta')} d\theta',
\end{equation}
where $R(r)$ and $\Theta(\theta)$ are functions of the radial and polar coordinates, respectively. This decomposition reveals the existence of a hidden symmetry captured by the generalized Carter constant
\begin{equation}
O = \mathscr{L} - (L_z - aE)^2,
\end{equation}
with $\mathscr{L}$ denoting the Carter separation constant. This constant is essential for the complete integrability of photon motion in the radial-polar $(r,\theta)$ plane. Explicitly, the equations of motion in this plane read
\begin{equation}
\begin{split}
\Sigma^2 \dot{t} &= \frac{\Xi^2}{\Delta_\theta} (L_z - a E \sin^2 \theta) + \frac{\Xi^2 (r^2 + a^2)}{\Delta_r} \left( (r^2 + a^2) E - a L_z \right),\\
\Sigma^2 \dot{\phi} &= \frac{\Xi^2}{\Delta_\theta} (L_z \csc^2\theta - a E) + \frac{a \Xi^2}{\Delta_r} \left( (r^2 + a^2) E - a L_z \right),\\
\Sigma^2 \dot{r}^2 &= R(r) = \Xi^2 \left( (r^2 + a^2) E - a L_z \right)^2 - \Delta_r \left[ (L_z - a E)^2 + O \right],\\
\Sigma^2 \dot{\theta}^2 &= \Theta(\theta) = \Delta_\theta \left[ (L_z - aE)^2 + O \right] + \Xi^2 \cos^2\theta \left( a^2 E^2 - L_z^2 \csc^2 \theta \right).
\end{split}
\end{equation}
The photon trajectories are conveniently characterized in terms of the dimensionless impact parameters $\xi = L_z / E$ and $\eta = \mathscr{L} / E^2$, which encode the ratio of conserved quantities. The radial and polar functions then take the forms
\begin{equation}
\begin{split}
R(r) &= E^2\Xi^2 \left( (r^2 + a^2) - a \xi \right)^2 - \Delta_r \left[ (\xi - a)^2 + \eta \right], \\
\Theta(\theta) &= E^2\eta \Delta_\theta + E^2\Xi^2 \cos^2\theta \left( a^2 - \xi^2 \csc^2 \theta \right).  
\end{split} \label{eq-R}
\end{equation}
Spherical photon orbits, which delineate the boundary of the black hole shadow, are determined by the conditions $R(r_p) = 0$ and $R'(r_p) = 0$ at the radius $r_p$ \cite{c30}. Solving Eq.~\eqref{eq-R} under these constraints yields
\begin{equation}
\begin{split}
\xi_p & = \frac{\Delta_r'(r_p^2 + a^2) - 4 r_p \Delta_r}{a \Delta_r'}, \\
\eta_p & = \frac{16 r_p^2 \Delta_r (a^2 - \Delta_r) - r_p^4 (\Delta_r')^2 + 8 r_p^3 \Delta_r \Delta_r'}{a^2 (\Delta_r')^2}.
\end{split}
\end{equation}
The celestial coordinates $(X, Y)$ for an observer situated at $r_o$, in the limit $r_o \to 10\,M$, and with an inclination angle $\theta_o$, are defined in a locally nonrotating reference frame as \cite{c35}
\begin{equation}
\begin{split}
X &= - r_o \frac{N(r_o)}{\sqrt{g_{\phi\phi}(r_o)}} \frac{\xi_p}{1 + \omega(r_o) \xi_p}, \\
Y & = \pm r_o \frac{N(r_o)}{\sqrt{g_{\theta\theta}(r_o)}} \frac{\sqrt{\Theta(\theta_o)}}{1 + \omega(r_o) \xi_p},
\end{split}
\end{equation}
where
\begin{equation}
\begin{split}
N(r) & = \sqrt{-\left( g_{tt}(r) + 2 \omega(r) g_{t\phi}(r) + \omega^2(r) g_{\phi\phi}(r) \right)}, \\
\omega(r) & = - \frac{g_{t\phi}(r)}{g_{\phi\phi}(r)}.
\end{split}
\end{equation}

\begin{figure*}[htbp]
	\centering
	
	\begin{subfigure}{0.45\textwidth}
		\centering
		\includegraphics[width=\linewidth]{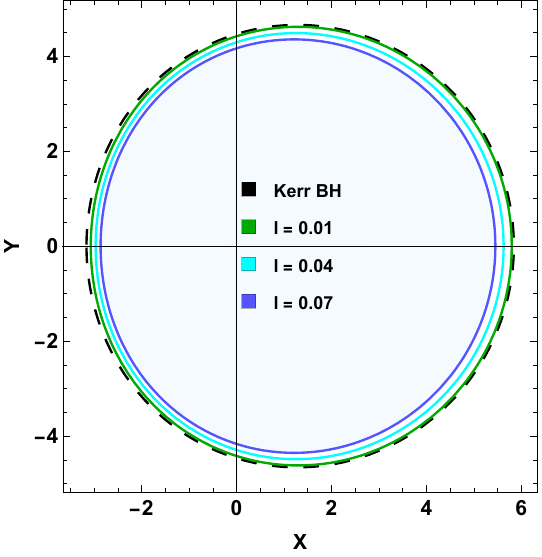}
		\caption{}
	\end{subfigure}
	\hfill
	\begin{subfigure}{0.45\textwidth}
		\centering
		\includegraphics[width=\linewidth]{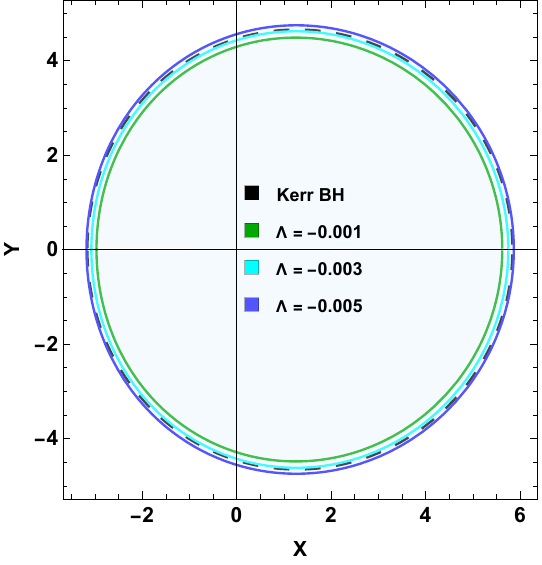}
		\caption{}
	\end{subfigure}
	
	\vspace{0.6em}
	
	\begin{subfigure}{0.45\textwidth}
		\centering
		\includegraphics[width=\linewidth]{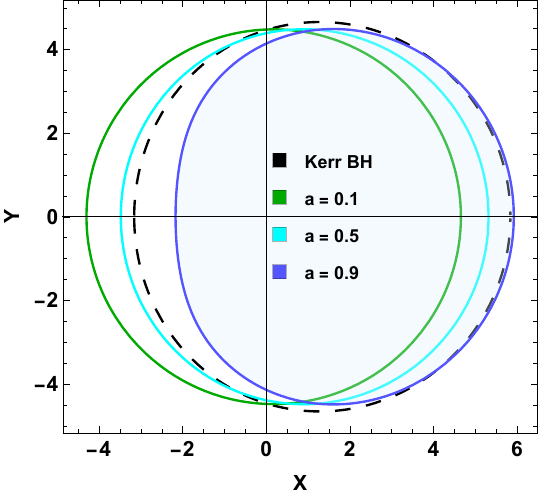}
		\caption{}
	\end{subfigure}
	\hfill
	\begin{subfigure}{0.45\textwidth}
		\centering
		\includegraphics[width=\linewidth]{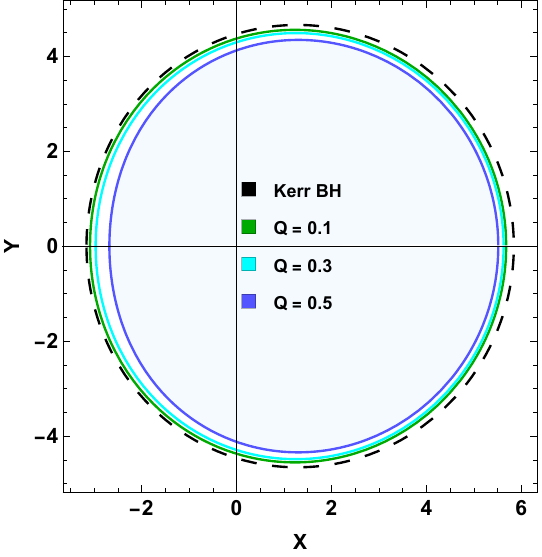}
		\caption{}
	\end{subfigure}
	
	\caption{Black hole shadow silhouettes for different values of the black hole parameters: (a) $l$ variation; (b) $\Lambda$ variation; (c) $a$ variation; (d) $Q$ variation.}
	\label{f-shadow}
\end{figure*}

The black hole shadow for an observer in the equatorial plane, $\theta_o = \pi/2$, is displayed in Fig.~\ref{f-shadow} for various combinations of spin, charge, cosmological constant, and Lorentz-violating parameter. As expected, increasing the rotation $a$ introduces asymmetry due to frame dragging, while larger charges $Q$ reduce the shadow size. The Lorentz-violating parameter $l$ subtly deforms the shadow, reflecting modifications of the effective gravitational potential in the photon sphere.  

\begin{figure}[htbp]
	\centering
	
	\begin{subfigure}{0.45\textwidth}
		\centering
		\includegraphics[width=\linewidth]{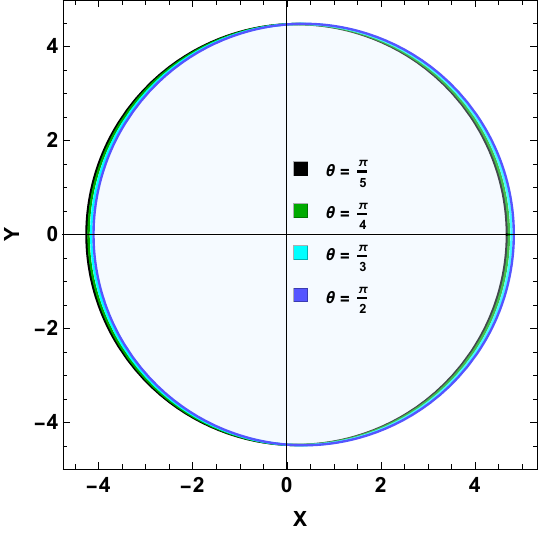}
		\caption{}
	\end{subfigure}
	\hfill
	\begin{subfigure}{0.45\textwidth}
		\centering
		\includegraphics[width=\linewidth]{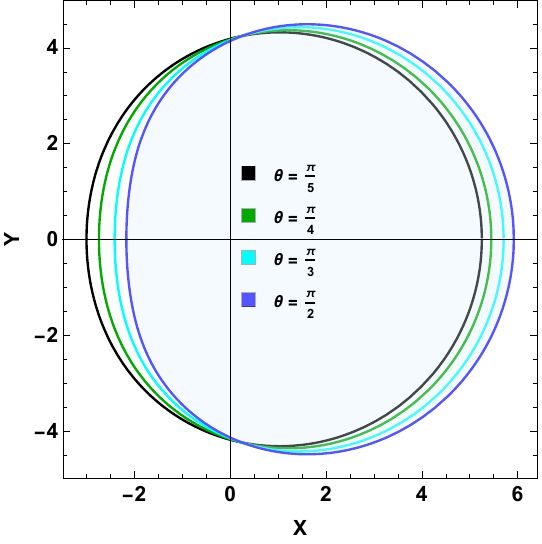}
		\caption{}
	\end{subfigure}
	
	\caption{Shadow silhouettes for different observer inclination angles $\theta$: (a) $a=0.2$, (b) $a=0.9$.}
	\label{f-shadow2}
\end{figure}

The effect of the observer inclination is illustrated in Fig.~\ref{f-shadow2}, showing that for higher spins, the asymmetry in the shadow becomes more pronounced when viewed off-axis. This demonstrates the importance of the $r$-$\theta$ coordinate dependence in accurately determining the apparent shadow shape.

Quantitative shadow observables, such as the linear radius $R_{sh}$, distortion $\delta_s$, area $A_s$, and angular diameter $\theta_{sh}$, provide a robust framework for connecting theoretical predictions with observations. The linear radius and distortion are defined as \cite{c25}
\begin{equation}
R_{sh} = \frac{(X_r - X_t)^2 + Y_t^2}{2 |X_r - X_t|}, \qquad
\delta_s = \frac{d_{cs}}{R_{sh}},
\end{equation}
where $d_{cs}$ represents the deviation of the shadow contour from a reference circle. The shadow area, which captures the overall extent of photon capture, is expressed as
\begin{equation}
A_s = 2 \int_{r_p^-}^{r_p^+} Y(r_p) X'(r_p) \, dr_p.
\end{equation}

\begin{figure*}[htbp]
	\centering
	
	\begin{subfigure}{0.45\textwidth}
		\centering
		\includegraphics[width=\linewidth]{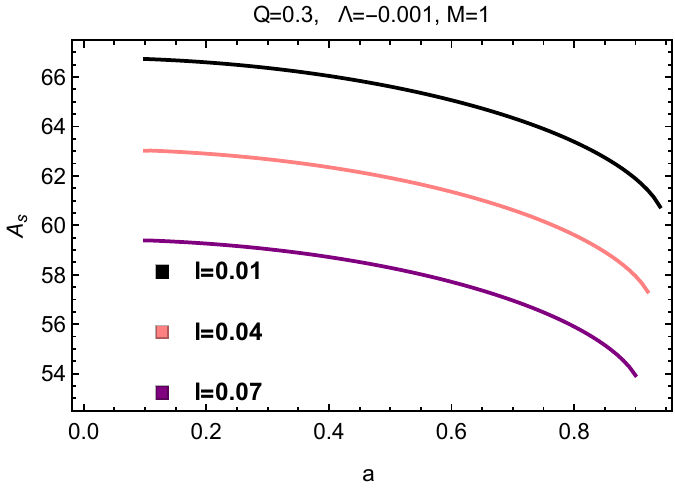}
		\caption{}
	\end{subfigure}
	\hfill
	\begin{subfigure}{0.45\textwidth}
		\centering
		\includegraphics[width=\linewidth]{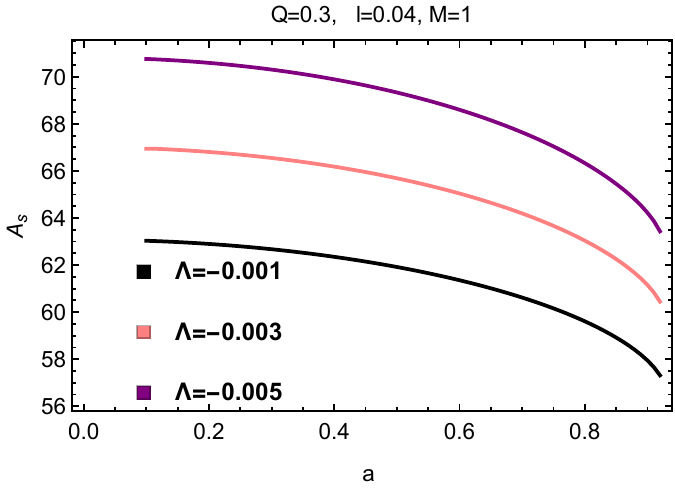}
		\caption{}
	\end{subfigure}
	\hfill
	\begin{subfigure}{0.45\textwidth}
		\centering
		\includegraphics[width=\linewidth]{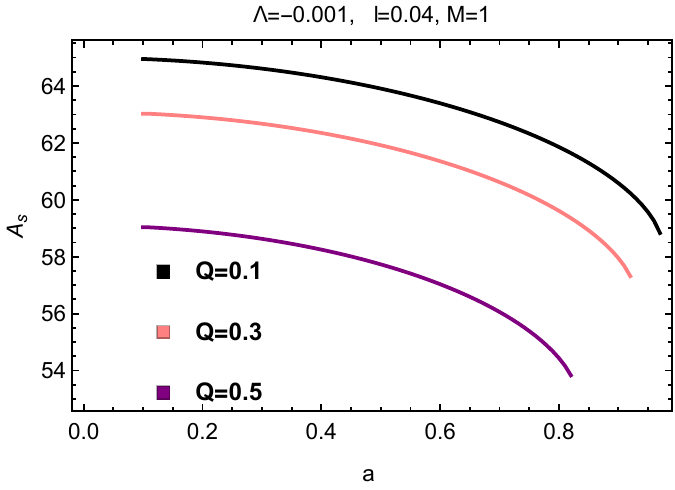}
		\caption{}
	\end{subfigure}
	
	\caption{Shadow area $A_s$ versus spin $a$ for varying Lorentz-violating parameter $l$, cosmological constant $\Lambda$, and electric charge $Q$.}
	\label{f-area}
\end{figure*}

Figure~\ref{f-area} illustrates how $A_s$ varies with the black hole spin $a$ under different $l$, $\Lambda$, and $Q$ values. As spin increases, the shadow area decreases, consistent with stronger frame-dragging effects that draw photon orbits closer to the horizon. Increasing $l$ modifies the gravitational potential and can slightly compress the shadow, while larger $Q$ decreases the photon capture cross-section. The negative cosmological constant $\Lambda$ focuses photon trajectories inward, partially counteracting the charge-induced expansion.

\begin{figure*}[htbp]
	\centering
	
	\begin{subfigure}{0.45\textwidth}
		\centering
		\includegraphics[width=\linewidth]{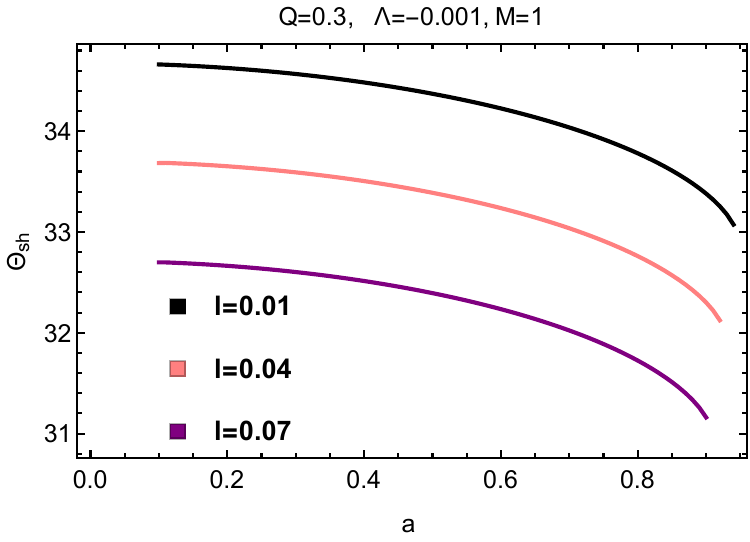}
		\caption{}
	\end{subfigure}
	\hfill
	\begin{subfigure}{0.45\textwidth}
		\centering
		\includegraphics[width=\linewidth]{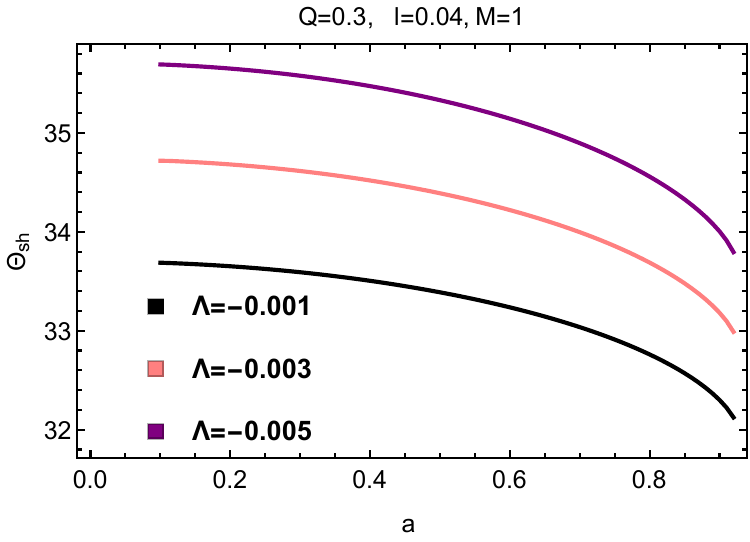}
		\caption{}
	\end{subfigure}
	
	\vspace{0.6em}
	
	\begin{subfigure}{0.45\textwidth}
		\centering
		\includegraphics[width=\linewidth]{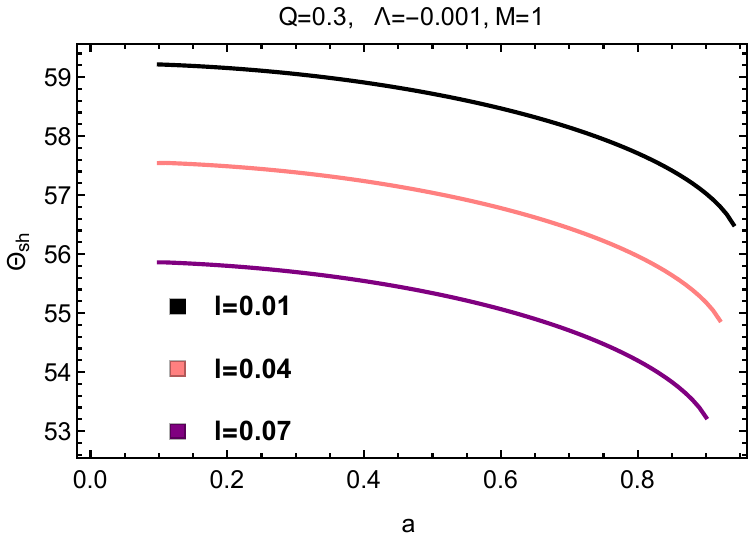}
		\caption{}
	\end{subfigure}
	\hfill
	\begin{subfigure}{0.45\textwidth}
		\centering
		\includegraphics[width=\linewidth]{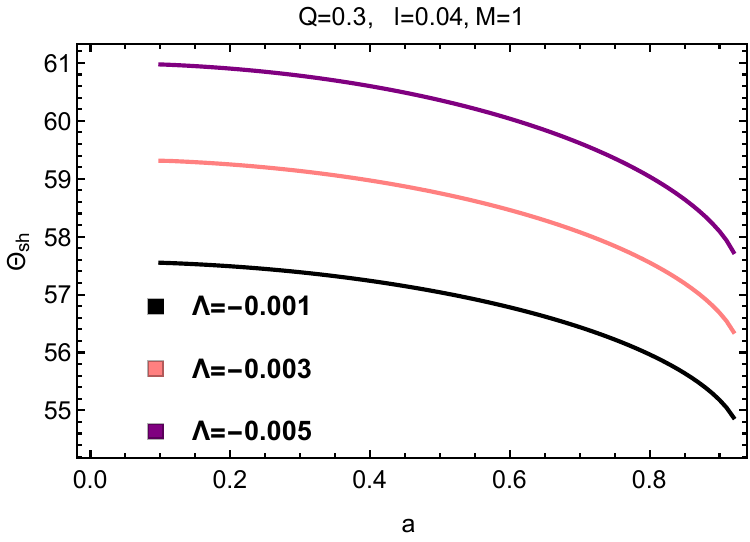}
		\caption{}
	\end{subfigure}
	
	\caption{Angular diameter $\theta_{sh}$ of the black hole shadow for different parameters: top panels correspond to M87$^*$, bottom panels to Sgr A$^*$ with representative values.}
	\label{f-theta}
\end{figure*}

The angular diameter of a black hole shadow represents one of the most direct observational signatures of the spacetime geometry in the vicinity of supermassive black holes (SMBHs), providing a crucial link between theoretical models and empirical measurements. For an observer located at a distance $D_0$ from the black hole, the angular size $\theta_{sh}$ of the shadow can be expressed in terms of the shadow radius $R_{sh}$ as \cite{c38,c39}:
\begin{equation}
2 R_{sh} = \frac{\theta_{sh} D_0}{M_{BH}},
\end{equation}
which, upon substitution and conversion to microarcseconds, yields
\begin{equation}
\theta_{sh} = 2 \times 9.87098 \times 10^{-6} \, R_{sh} \, \frac{M_{BH}}{M_\odot} \, \frac{1 \text{ kpc}}{D_0} \, \mu as,
\end{equation}
where $M_{BH}$ represents the mass of the black hole and $D_0$ is the distance from the observer to the black hole. This formulation explicitly connects the theoretical shadow radius, which depends on the underlying geometry and black hole parameters, to a directly measurable angular quantity. To illustrate, for the supermassive black holes (SMBHs) at the center of the M87 galaxy, we adopt $M_{M87^*} = 6.4 \times 10^9 M_\odot$ and a distance $D_0 = 16.8$ Mpc, while for Sgr A$^*$, the supermassive black hole (SMBH) at the center of the Milky Way, we take $M_{Sgr A^*} = 4.1 \times 10^6 M_\odot$ and $D_0 = 8.3$ kpc. Using these reference values, the resulting angular diameters of the shadow can be computed for a wide range of black hole parameters, including the dimensionless spin $a$, electric charge $Q$, Lorentz-violating parameter $l$, and the cosmological constant $\Lambda$. These computations, plotted in Fig.~\ref{f-theta}, allow a detailed visualization of how variations in each parameter individually, as well as in combination, influence the apparent size and asymmetry of the shadow. 

The spin parameter $a$ primarily introduces asymmetry and D-shaped distortions due to frame-dragging effects: as the spin increases, the shadow becomes progressively flattened on the side co-rotating with the black hole, shifting the centroid of the shadow and altering $R_{sh}$. The electric charge $Q$ modifies the effective gravitational potential, generally reducing the photon capture radius and inducing a mild lateral compression of the shadow. The Lorentz-violating parameter $l$, characteristic of Bumblebee gravity, affects the spacetime structure near the horizon, elongating the shadow horizontally and enhancing asymmetry. Lastly, the negative cosmological constant $\Lambda$ corresponding to the Anti-de Sitter (AdS) curvature introduces a focusing effect on photon trajectories, effectively increasing the vertical extension of the shadow while interacting subtly with the other parameters. 

Taken together, these results demonstrate that the angular diameter $\theta_{sh}$ is a sensitive probe of the fundamental black hole parameters. By comparing the calculated angular diameters and shadow shapes with high-resolution Event Horizon Telescope (EHT) observations of M87$^*$ and Sgr A$^*$, it is possible to place stringent constraints on the spin, charge, Lorentz-violating effects, and cosmological curvature associated with Bumblebee Kerr-Newman-AdS black holes. Such analyses not only serve as a test of alternative gravity models but also provide insight into the photon capture region and the near-horizon geometry, thereby bridging theoretical predictions and astrophysical observations with unprecedented precision.

\begin{figure}[t]
\centering

\includegraphics[width=.40\textwidth]{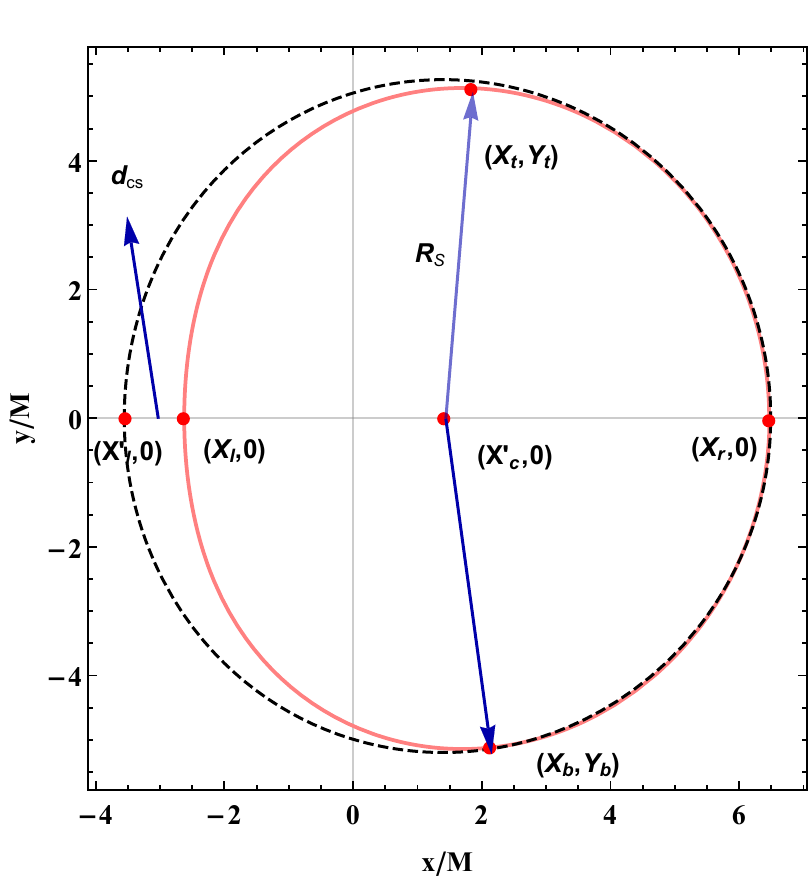}

\caption{Schematic representation of shadow observables: the shadow radius $R_s$, distortion parameter $\delta_s$, and the shadow center $X'_c$, which together characterize the size and asymmetry of the Bumblebee Kerr--Newman--AdS black hole shadow.}
\label{shadow}
\end{figure}

Finally, the detailed shadow parameters for various combinations of the spin $a$, electric charge $Q$, cosmological constant $\Lambda$, and Lorentz-violating parameter $l$ are systematically summarized in Table~\ref{tab2}. The shadow center, denoted as $X'_c$, has been calculated using two complementary approaches: the averaging method (AM) and the integration method (IM). In the integration method, $X'_c$ is determined by \cite{BumblebeeEHT,Hioki2009}:
\begin{equation}
X'_c = \frac{2}{A_s} \int_{r_p^-}^{r_p^+} Y(r_p) X(r_p) X'(r_p) \, dr_p,
\end{equation}
where $A_s$ represents the shadow area, and $r_p^\pm$ correspond to the extremal radii of the photon sphere. In contrast, the averaging method provides a simplified estimate:
\begin{equation}
X'_c = \frac{X_r - |X_l|}{2},
\end{equation}
where $X_r$ and $X_l$ denote the rightmost and leftmost points of the shadow, respectively. The shadow geometry, including $X_r$, $X_l$, $X_t$, $Y_t$, $X_b$, and $Y_b$, along with the radius $R_s$ and the distortion parameter $\delta_s$, which together quantify the apparent size and asymmetry of the shadow, is illustrated in Fig.~\ref{shadow}.

\begin{table*}[t]
\centering
\captionsetup{font=small, labelfont=bf}
\caption{Shadow parameters for various values of spin, charge, cosmological constant, and Lorentz-violating parameter. $X'_c$ is computed via averaging (AM) and integration (IM) methods.}

\renewcommand{\arraystretch}{1.1}
\setlength{\tabcolsep}{3pt}
\rowcolors{3}{lightgray}{white}
\resizebox{0.98\textwidth}{!}{%
\begin{tabular}{lccccccccc}
\rowcolor{headerblue}
\toprule
\textbf{Parameters} & \textbf{$X_r$} & \textbf{$X_l$} & \textbf{$X_t$} & \textbf{$Y_t$} & \textbf{$X_b$} & \textbf{$Y_b$} & \textbf{$X'_c$(AM)} & \textbf{$X'_c$(IM)} & \textbf{$X'_l$} \\
\midrule
$\Lambda=-0.001$, $a=0.7$, $l=0.01$, $Q=0.3$ & 5.7864 & -3.0697 & 1.2858 & 4.6155 & 1.2858 & -4.6155 & 1.0447 & 1.3378 & -4.6104 \\
$\Lambda=-0.001$, $a=0.7$, $l=0.04$, $Q=0.3$ & 5.6181 & -2.9665 & 1.2495 & 4.4846 & 1.2495 & -4.4846 & 1.2265 & 1.3041 & -4.4809 \\
$\Lambda=-0.001$, $a=0.7$, $l=0.07$, $Q=0.3$ & 5.4476 & -2.8612 & 1.2125 & 4.3521 & 1.2125 & -4.3521 & 1.2932 & 1.2701 & -4.3496 \\
$\Lambda=-0.003$, $a=0.7$, $l=0.04$, $Q=0.3$ & 5.7507 & -3.0734 & 1.2593 & 4.6201 & 1.2593 & -4.6201 & 1.3386 & 1.3161 & -4.6183 \\
$\Lambda=-0.005$, $a=0.7$, $l=0.04$, $Q=0.3$ & 5.8723 & -3.1756 & 1.2660 & 4.7475 & 1.2660 & -4.7475 & 1.3483 & 1.3250 & -4.8698 \\
$\Lambda=-0.001$, $a=0.1$, $l=0.04$, $Q=0.3$ & 4.6552 & -4.3001 & 0.1774 & 4.4809 & 0.1775 & -4.4809 & 0.1774 & 0.1775 & -4.4809 \\
$\Lambda=-0.001$, $a=0.5$, $l=0.04$, $Q=0.3$ & 5.3090 & -3.4810 & 0.8910 & 4.4822 & 0.8910 & -4.4822 & 0.9140 & 0.9080 & -4.4809 \\
$\Lambda=-0.001$, $a=0.9$, $l=0.04$, $Q=0.3$ & 5.9189 & -2.1684 & 1.6084 & 4.4888 & 1.6084 & -4.4888 & 1.8752 & 1.7817 & -4.4809 \\
$\Lambda=-0.001$, $a=0.7$, $l=0.04$, $Q=0.1$ & 5.6716 & -3.0848 & 1.2289 & 4.5521 & 1.2289 & -4.5521 & 1.2934 & 1.2753 & -4.5483 \\
$\Lambda=-0.001$, $a=0.7$, $l=0.04$, $Q=0.5$ & 5.5063 & -2.6764 & 1.2983 & 4.3407 & 1.2983 & -4.3407 & 1.4149 & 1.3799 & -4.3371 \\
\bottomrule
\end{tabular}
} 
\label{tab2}
\end{table*}

A detailed inspection of Table~\ref{tab2} shows that the Lorentz-violating parameter $l$ induces a progressive horizontal shift of the shadow center $X'_c$ toward larger values, reflecting subtle asymmetry along the $X$-axis, while also causing minor elongation. Increasing the spin $a$ enhances the characteristic D-shaped distortion, as seen in the asymmetric displacement of $X_r$ and $X_l$, which results from frame-dragging effects on photon trajectories. The electric charge $Q$ reduces the photon capture radius, producing a lateral compression of the shadow, whereas a negative cosmological constant $\Lambda$ slightly stretches the shadow vertically by focusing photon paths in AdS spacetime. Comparison between the averaging and integration methods shows that the latter consistently yields larger $X'_c$ values for asymmetric configurations, demonstrating its sensitivity to detailed photon distribution. Collectively, these results provide a precise and quantitative characterization of the shadow’s size, shape, and asymmetry, offering valuable information for constraining the physical parameters of Bumblebee Kerr-Newman-AdS black holes and for linking theoretical predictions with high-resolution observations such as those by the Event Horizon Telescope.

\subsection{Ergosphere of Bumblebee Kerr-Newman-AdS Black Holes} \label{subsec:ergosphere}

The ergosphere of a rotating Bumblebee Kerr-Newman-AdS black hole is a distinct region of spacetime located outside the event horizon, in which frame-dragging effects are so intense that no object can remain stationary relative to a distant inertial observer. This phenomenon originates from the rotation of the black hole, quantified by the spin parameter $a$, which, according to general relativity, drags spacetime along its rotation axis, compelling all matter, particles, and photons to co-rotate. The ergosphere is bounded internally by the event horizon at radius $r_h$, determined by $g_{rr} = \Delta_r = 0$, and externally by the static limit at radius $r_s$, defined where the timelike Killing vector becomes null, $g_{tt} = 0$. At the poles of the black hole, the horizon and static limit are tangent, while at the equatorial plane the static limit extends outward beyond the horizon, producing the characteristic oblate, peanut-shaped geometry. The additional parameters, including the Lorentz symmetry-violating factor $l$, the electric charge $Q$, and the cosmological constant $\Lambda$, further deform the ergosphere, introducing anisotropic modifications to the spacetime and altering the energy dynamics within the black hole environment. The ergosphere can therefore be regarded as a fundamental region of the spacetime geometry surrounding a rotating black hole, essential for its overall dynamics, and its study is crucial for understanding mechanisms such as relativistic jets, high-energy particle acceleration, and energy extraction processes \cite{c31,c32,c33}.

\begin{figure*}[htbp]
	\centering
	
	\begin{subfigure}{0.3\textwidth}
		\centering
		\includegraphics[width=\linewidth]{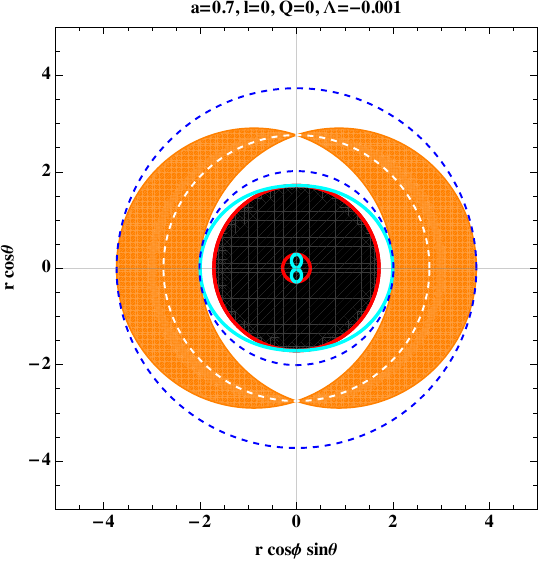}
		\caption{}
	\end{subfigure}
	\hfill
	\begin{subfigure}{0.3\textwidth}
		\centering
		\includegraphics[width=\linewidth]{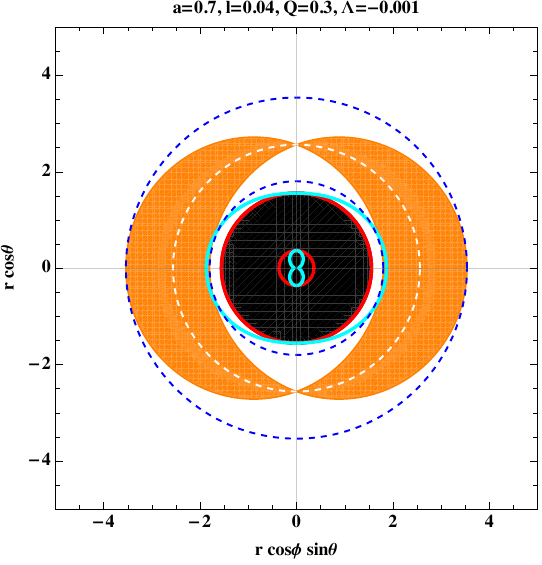}
		\caption{}
	\end{subfigure}
	\hfill
	\begin{subfigure}{0.3\textwidth}
		\centering
		\includegraphics[width=\linewidth]{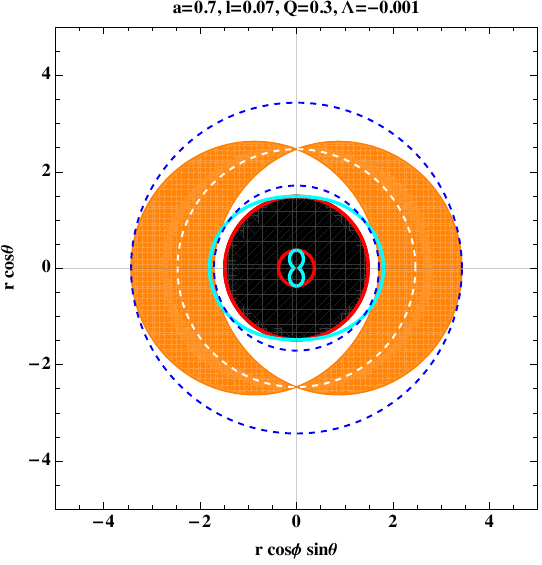}
		\caption{}
	\end{subfigure}
	
	\vspace{0.6em}
	
	\begin{subfigure}{0.3\textwidth}
		\centering
		\includegraphics[width=\linewidth]{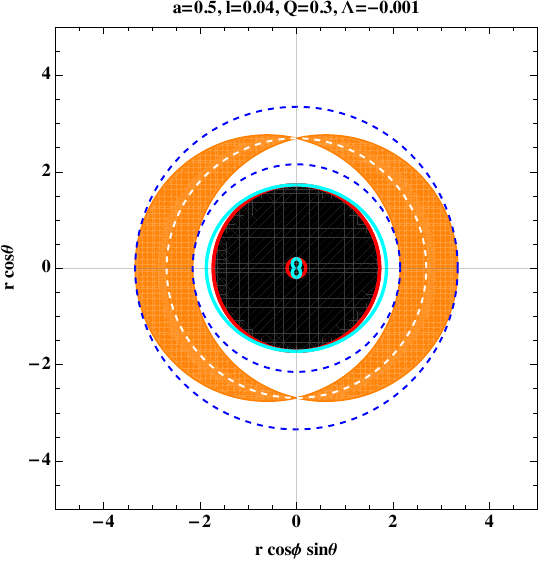}
		\caption{}
	\end{subfigure}
	\hfill
	\begin{subfigure}{0.3\textwidth}
		\centering
		\includegraphics[width=\linewidth]{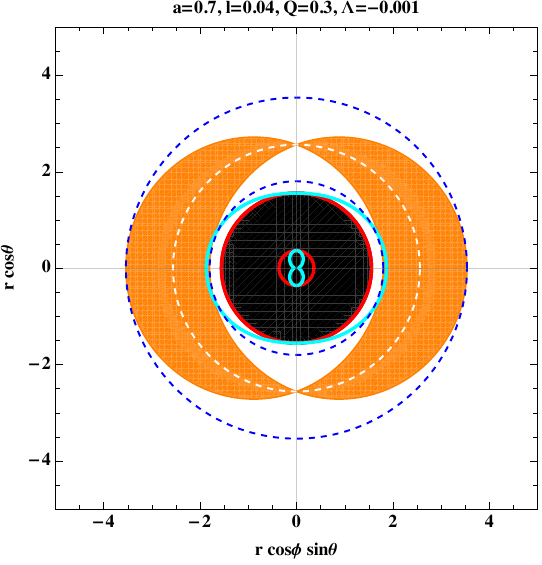}
		\caption{}
	\end{subfigure}
	\hfill
	\begin{subfigure}{0.3\textwidth}
		\centering
		\includegraphics[width=\linewidth]{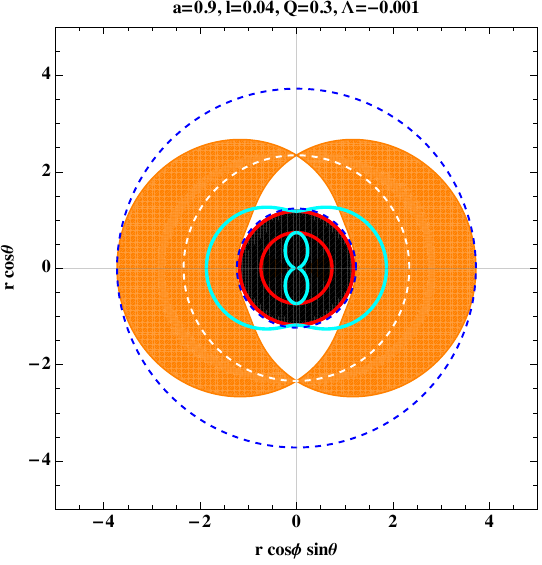}
		\caption{}
	\end{subfigure}
	
	\vspace{0.6em}
	
	\begin{subfigure}{0.3\textwidth}
		\centering
		\includegraphics[width=\linewidth]{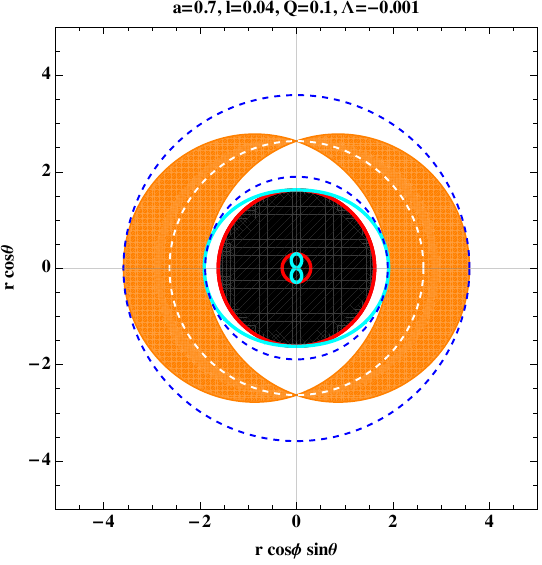}
		\caption{}
	\end{subfigure}
	\hfill
	\begin{subfigure}{0.3\textwidth}
		\centering
		\includegraphics[width=\linewidth]{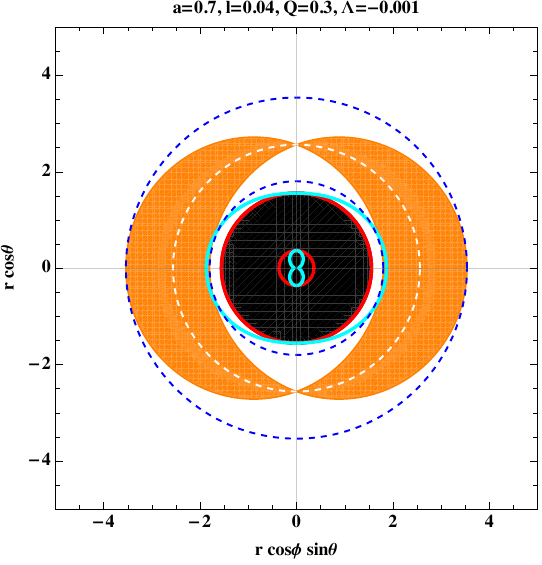}
		\caption{}
	\end{subfigure}
	\hfill
	\begin{subfigure}{0.3\textwidth}
		\centering
		\includegraphics[width=\linewidth]{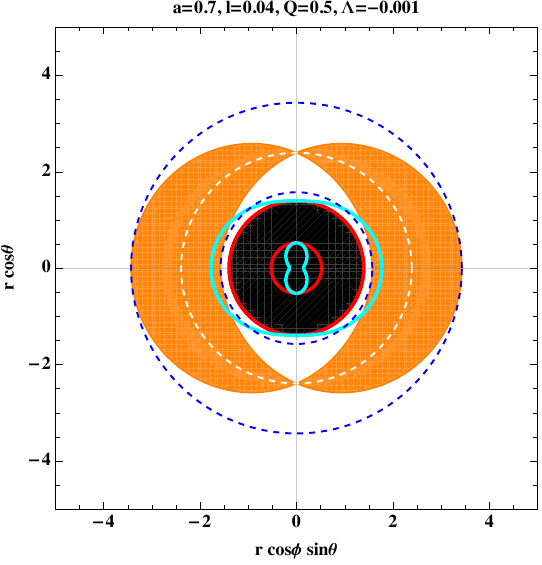}
		\caption{}
	\end{subfigure}
	
	\caption{The behavior of the ergosphere region in the $xz$ plane of the Bumblebee Kerr-Newman-AdS black hole spacetime. Blue and pink lines denote the static limit surface and event horizon, respectively. Variations in spin parameter $a$, Lorentz-violating parameter $l$, cosmological constant $\Lambda$, and electric charge $Q$ significantly modify the shape, size, and photon region of the ergosphere.}
	\label{f-3}
\end{figure*}

In a rotating black hole, the ergosphere lies between the event horizon $r_h$ where $g_{rr} = \Delta_r = 0$ and the static limit $r_s$ where $g_{tt} = 0$. Within this region, no observer can remain at rest relative to infinity, and the frame-dragging effect forces all particles and photons to co-rotate with the black hole. The photon region, encompassing unstable circular photon orbits, is embedded within the ergosphere and responds sensitively to variations in spin $a$, Lorentz-violating parameter $l$, electric charge $Q$, and cosmological constant $\Lambda$.

Panel (a) in Figure~\ref{f-3} corresponds to the standard Kerr geometry with $l = 0$ and $Q = 0$. For moderate rotation, the ergosphere exhibits the classical peanut-shaped configuration, which is axially symmetric and moderately oblate. The static limit and horizon coincide at the poles and reach maximum separation at the equatorial plane. Photon orbits form nearly circular, tightly bound rings around the equatorial plane, defining a thin photon region close to the horizon. 

Panel (b) shows the effect of introducing non-zero spin $a$ and electric charge $Q$. The static limit, indicated by the blue curve, shifts outward relative to the nearly stationary event horizon, represented by the pink curve. This displacement enlarges the ergosphere and expands the photon region. The Lorentz symmetry-violating background contributes anisotropic spacetime corrections that effectively increase the frame-dragging strength and the spatial domain available for extraction of rotational energy.

Panel (c) demonstrates the combined effect of a stronger Lorentz violation with $l = 0.07$. The static limit at the equator exhibits a pronounced asymmetric bulge. The photon region extends to larger radii, allowing photons to orbit further from the horizon. This deformation increases the volume of the ergosphere and enhances the potential for energy extraction through the Penrose process, in which a particle entering the ergosphere can split, with one fragment falling into the black hole and the other escaping with energy exceeding that of the initial particle. This process reduces the black hole's angular momentum.

Panels (d) through (f) depict variations in the spin parameter $a$. For lower spin $a = 0.5$, the ergosphere contracts and the static limit and event horizon nearly coincide, reducing the photon trapping region and the available rotational energy. For higher spin $a = 0.9$, the equatorial separation increases substantially, producing a more oblate photon region and increasing the strength of frame-dragging, thereby enlarging the ergosphere volume and the potential for energy extraction.

Panels (g) through (i) illustrate the impact of increasing Lorentz violation from $l = 0.1$ to $l = 0.5$. The effective gravitational pull is reduced, pushing both the static limit and event horizon outward. The presence of electric charge $Q$ counteracts gravity, causing the event horizon radius to decrease more rapidly than the static limit radius, thereby widening the ergosphere. Photon orbits become less tightly bound, extend to larger radii, and are increasingly sensitive to asymmetries. Positive cosmological constant $\Lambda$ amplifies this outward expansion, whereas negative $\Lambda$ slightly mitigates it.

The event horizon is at $\Delta_r(r_h)=0$ (we already use this), while the static limit surface (outer boundary of the ergosphere) is defined by
\begin{equation}\label{eq-E1}
g_{tt}(r_{s},\theta)=0 \quad \Longrightarrow \quad \Delta_{r}(r_{s}) = a^{2}\Delta_{\theta}(r_{s},\theta)\sin^{2}\theta .
\end{equation}
This is the exact condition for the static limit in the Bumblebee Kerr-Newman-AdS spacetime. On the equatorial plane $\theta=\pi/2$. 
Then Eq. \eqref{eq-E1} reduces to the simple condition
\begin{equation}\label{eq-E2}
\Delta_{r}(r_{s}^{\rm eq}) = a^{2}. 
\end{equation}
Inserting your explicit $\Delta_r$ and subtracting $a^2$, one finds
\begin{equation}
\Delta_{r} - a^{2} = r^{2} - 2Mr + \frac{Q^{2}}{(1-l)^{2}}- \frac{\Lambda r^{2}}{3(1-l)}(r^{2}+a^{2})+ \frac{l}{1-l} r^{2}.
\end{equation}
Exactly as in this horizon analysis, this can be written as a depressed quartic
\begin{equation}\label{eq-E3}
\alpha r^{4} + b r^{2} - 2Mr + c_{s} = 0,
\end{equation}
with the same coefficients $\alpha$ and $b$ as in Eqs. \eqref{e-2-5}-\eqref{e-2-6}, but with a different constant term $c_{s} = \frac{Q^{2}}{(1-l)^{2}}$ evaluated at $r_{s}^{eq}$. For comparison, our horizon quartic has $	c_{h} = a^{2} + \frac{Q^{2}}{(1-l)^{2}}$ evaluated at $r_{h}$. So the equatorial static limit radius $r_s^{(eq)}$ is the largest real root of Eq. \eqref{eq-E3}, whereas the event horizon $r_+$ is the largest real root of
\begin{equation}\label{eq-5-22}
\alpha r^{4} + b r^{2} - 2Mr + c_{h} = 0.
\end{equation}
Equations \eqref{eq-E3} and \eqref{eq-5-22} are similar, and both are zero at the ergosphere surface and the event horizon, respectively. Therefore, we can write a general form for them as:
\begin{equation}
G(r;c)=\alpha r^{4} + b r^{2} - 2Mr + c,
\end{equation}
so that $G(r_{+};c_{h})=0$ and $ G(r_{s}^{\rm eq};c_{s})=0$ with $c_{h} = c_{s}+a^{2}$. Treating $c$ as a parameter, the shift of a root under a change $\delta c$ is
\begin{equation}
	\frac{dr}{dc}
	= -\frac{\partial G/\partial c}{\partial G/\partial r}
	= -\frac{1}{4\alpha r^{3}+2br-2M}.
\end{equation}
To go from the horizon (with $c=c_h$) to the equatorial static limit (with $c=c_s=c_h - a^2$), we take $ \delta c = -a^2$.
Evaluating the derivative at $r=r_+$ gives, to leading order in $a^2$,
\begin{equation}\label{eq-E4}
	\Delta r_{  eq}
	\equiv r_{s}^{ (eq)}-r_{+}
	\simeq \frac{a^{2}}{4\alpha r_{+}^{3}+2 b r_{+}-2M}.
\end{equation}
For an AdS background with $\Lambda<0 \rightarrow \alpha>0$, and for typical parameter ranges, the denominator is positive, so $\Delta r_{eq}>0$ as expected: the static limit lies outside the horizon at the equator. We can see that the Lorentz-violating parameter $l$ and the charge $Q$ enter Eq. \eqref{eq-E4} through both $\alpha$ and $b$ and through the value of $r_+$.This expression captures the three-dimensional volume enclosed between the horizon and the static limit surface:
\begin{equation}\label{e-5-22}
V_{\rm erg} \approx 2 \pi \int_0^\pi \sin\theta \int_{r_h}^{r_s(\theta)} r^2 \, dr \, d\theta,
\end{equation}
where $r_h$ denotes the event horizon radius and $r_s(\theta)$ denotes the static limit radius at polar angle $\theta$. In a slow-rotation approximation one may model the angular dependence as
\begin{equation}\label{eq-E5}
	r_{h}(\theta)\simeq r_{+}, \qquad
	r_{s}(\theta)\simeq r_{+}+\Delta r_{\rm eq}\sin^{2}\theta ,
\end{equation}
i.e., the separation is maximal at the equator and vanishes at the poles, as in the Kerr limit. Inserting Eq. \eqref{eq-E5} into \eqref{e-5-22} and expanding to first order in $\Delta r_{eq} \ll r_+$ ,
\begin{equation}\label{eq-E6}
	\begin{split}
		V_{erg} \simeq &2\pi \int_0^\pi \sin\theta \int_{r_+}^{r_+ + \Delta r_{eq}\sin^2\theta} r^2 dr\, d\theta\\
		= &2\pi \int_0^\pi \sin\theta \left[ \frac{(r_+ + \Delta r_{eq}\sin^2\theta )^3 - r_+^3 }{3}\right] d\theta\\
		\simeq& 2\pi \int_0^\pi \sin\theta \left[ r_+^2 \Delta r_{eq}\sin^2\theta \right] d\theta\\
		= &2\pi r_+^2 \Delta r_{eq} \int_0^\pi \sin^3\theta\, d\theta
		= \frac{8\pi}{3} r_+^2 \Delta r_{eq} .
	\end{split} 
\end{equation}
Using Eq. \eqref{eq-E4},
\begin{equation}\label{eq-E7}
	V_{erg}
	\simeq \frac{8\pi}{3} r_{+}^{2}
	\frac{a^{2}}{4\alpha r_{+}^{3}+2 b r_{+}-2M}.
\end{equation}
This gives us a fully analytic first-order expression for the volume of the ergosphere in terms of $a$, $Q$, $l$, $\Lambda$, $M$. We can then say that in the limit $\Lambda\to 0$, $l\to 0$, this reproduces the familiar Kerr scaling $V_{erg} \propto a^2 M^2 $ \cite{c31,c32,c33}.

\section{Penrose Extraction with Lorentz-Symmetry Breaking}\label{sec:Penrose}

In this section, we analyze the effect of Lorentz-symmetry breaking on the Penrose energy extraction mechanism in the Bumblebee Kerr-Newman-AdS spacetime. Lorentz violation is characterized by the dimensionless Bumblebee parameter $l$, which modifies both the spacetime geometry and the rotational properties of the black hole \cite{liu2025charged,b10}. Our aim is to determine how this parameter alters the structure of the ergoregion, the availability of negative-energy states, and the maximum efficiency of rotational energy extraction. The Bumblebee Kerr-Newman-AdS spacetime is stationary and axisymmetric, admitting the Killing vectors
\begin{equation}
\xi^\mu_{(t)}=\partial_t,
\qquad
\xi^\mu_{(\phi)}=\partial_\phi .
\end{equation}
These symmetries guarantee the conservation of energy and azimuthal angular momentum along geodesic motion \cite{c29}. For a test particle of rest mass $m$, four-velocity $u^\mu=dx^\mu/d\tau$, and four-momentum $p^\mu=mu^\mu$, the conserved quantities measured at infinity are defined covariantly as
\begin{equation}
E=-p_\mu \xi^\mu_{(t)}=-p_t,
\qquad
L=p_\mu \xi^\mu_{(\phi)}=p_\phi .
\end{equation}
In terms of the metric components, these quantities take the explicit form \cite{aa-7,a,c30}
\begin{equation}
E=-g_{tt}\dot t-g_{t\phi}\dot\phi,
\qquad
L=g_{t\phi}\dot t+g_{\phi\phi}\dot\phi .
\end{equation}
The event horizon is located at the largest root of the radial function \cite{a}
\begin{equation}
\Delta_r(r)=0,
\end{equation}
which depends on the Lorentz-violating parameter $l$, the electric charge $Q$, the rotation parameter $a$, and the cosmological constant $\Lambda$ \cite{liu2025charged}. The static limit surface is defined by
\begin{equation}
g_{tt}(r_s,\theta)=0 .
\end{equation}
The region bounded by
\begin{equation}
r_h<r<r_s(\theta)
\end{equation}
defines the ergoregion. Within this region, the timelike Killing vector $\xi^\mu_{(t)}$ becomes spacelike ($g_{tt}>0$), permitting the conserved energy $E=-p_t$ to assume negative values \cite{aa-5}. For fixed $(a,Q,\Lambda)$ and within the physically admissible parameter range of the Bumblebee Kerr-Newman-AdS solution, the Lorentz-violating parameter modifies the metric functions such that the static limit surface expands outward more rapidly than the event horizon. This behavior can be expressed as
\begin{equation}
\frac{\partial r_s}{\partial l}>0,
\qquad
\frac{\partial r_h}{\partial l}<\frac{\partial r_s}{\partial l},
\end{equation}
demonstrating that the ergoregion grows as $l$ increases \cite{liu2025charged,b10}. This geometric enlargement increases the domain in which negative-energy trajectories are allowed. All massive particles satisfy the normalization condition
\begin{equation}
g_{\mu\nu}u^\mu u^\nu=-1 .
\end{equation}
Restricting to equatorial motion $(\theta=\pi/2)$ gives
\begin{equation}
g_{tt}\dot t^2+2g_{t\phi}\dot t\dot\phi+g_{\phi\phi}\dot\phi^2+g_{rr}\dot r^2=-1 .
\end{equation}
Solving the system defined by the conserved quantities $E$ and $L$ for $\dot t$ and $\dot\phi$ yields \cite{a,c30}
\begin{equation}
\dot t=\frac{g_{\phi\phi}E+g_{t\phi}L}{g_{t\phi}^2-g_{tt}g_{\phi\phi}},
\qquad
\dot\phi=\frac{-g_{t\phi}E-g_{tt}L}{g_{t\phi}^2-g_{tt}g_{\phi\phi}} .
\end{equation}
Introducing the frame-dragging angular velocity
\begin{equation}
\Omega=-\frac{g_{t\phi}}{g_{\phi\phi}},
\end{equation}
the condition for negative-energy motion becomes \cite{ishii2022energy}
\begin{equation}
E<\Omega L .
\end{equation}
Since both $g_{t\phi}$ and $g_{\phi\phi}$ depend explicitly on $l$, the allowed phase space of negative-energy trajectories is modified accordingly \cite{Wald1984,aa-7}. For fixed $(a,Q,\Lambda)$, the frame-dragging angular velocity increases monotonically with $l$, lowering the threshold for negative-energy motion. The Penrose process occurs when a particle entering the ergoregion undergoes the decay
\begin{equation}
(0)\rightarrow(1)+(2),
\end{equation}
subject to the conservation laws
\begin{equation}
E_0=E_1+E_2,
\qquad
L_0=L_1+L_2 .
\end{equation}
If fragment $(1)$ falls into the black hole with negative energy $(E_1<0)$, the escaping fragment satisfies
\begin{equation}
E_2=E_0-E_1>E_0 ,
\end{equation}
demonstrating net extraction of energy from the black hole \cite{Penrose1969,aa-7}. The maximum efficiency is achieved when the decay occurs arbitrarily close to the event horizon. The angular velocity of the horizon is defined by
\begin{equation}
\Omega_h=-\left.\frac{g_{t\phi}}{g_{\phi\phi}}\right|_{r=r_h} .
\end{equation}
For the Bumblebee Kerr-Newman-AdS spacetime, this quantity is
\begin{equation}\label{metric-functions}
\Omega_h=\frac{a\,\Xi}{r_h^2+a^2},
\qquad
\Xi=1+\frac{\Lambda a^2}{3(1-l)} .
\end{equation}
The explicit dependence on $l$ implies that $\Omega_h$ increases as Lorentz symmetry breaking strengthens \cite{Gibbons2005,liu2025charged}. The minimum energy of the infalling fragment is therefore
\begin{equation}
E_{1,\min}=\Omega_h L_1 ,
\end{equation}
becoming increasingly negative for larger $l$. Defining the Penrose efficiency \cite{penrose-nat} as
\begin{equation} 
	\eta = \frac{E_{\rm extracted}}{E_0} = \frac{E_2 - E_0}{E_0} = \frac{(E_0 - E_1) - E_0}{E_0} = -\frac{E_1}{E_0},
\end{equation}
the maximum efficiency, achieved for an optimal decay arbitrarily close to the horizon \cite{efficiency}, is
\begin{equation}
	\eta_{\max} = \frac{1}{2}\left(\frac{1}{\sqrt{1-\Omega_h(l) a}} - 1\right).
\end{equation}
For $\eta_{\max} > 0$, the necessary and sufficient conditions are compactly: the particle must lie inside the ergoregion, $r_h < r < r_s(\theta)$; the infalling fragment must carry negative energy, $E_1 =\Omega_h(l)\, L_1 < 0$, which for a prograde black hole ($\Omega_h(l) > 0$) requires the fragment to counter-rotate, $L_1 < 0$; the horizon angular velocity must be sub-extremal, $0 < \Omega_h(l)\, a < 1$, ensuring the efficiency is real; and the Lorentz-violating parameter must be physically admissible, $0 \le l < 1$. Under these conditions, energy extraction is realized with $E_2 = E_0 - E_1 > E_0$. In the limiting case of a non-rotating black hole ($a = 0$), the horizon angular velocity vanishes, $\Omega_h = 0$, yielding \(\eta_{\max} = 0\) demonstrating that the Penrose process is impossible without rotation. Consequently, since $\Omega_h(l)$ increases with the Lorentz-violating parameter $l$ (see Eq. \eqref{metric-functions}), the efficiency of rotational energy extraction is enhanced relative to the Lorentz-invariant Kerr-Newman-AdS spacetime \cite{Wagh1989,Schnittman2014}.

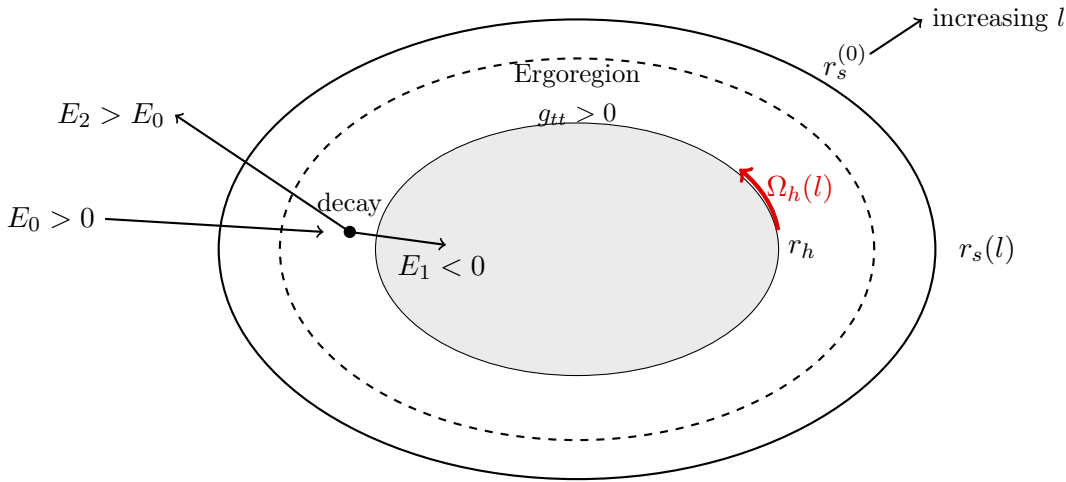
\begin{figure}[t]
\centering
\begin{tikzpicture}[scale=1.15]

\definecolor{ergocolor}{RGB}{235,235,235}
\definecolor{horizoncolor}{RGB}{185,185,185}
\definecolor{omegaColor}{RGB}{220,0,0} 

\fill[horizoncolor] (0,0) ellipse (2.3cm and 1.45cm);
\draw[thick] (0,0) ellipse (2.3cm and 1.45cm);
\node[right] at (2.3,0) {$r_h$};

\draw[thick,dashed] (0,0) ellipse (3.4cm and 2.2cm);
\node[above right] at (2.7,1.85) {$r_s^{(0)}$};

\draw[thick] (0,0) ellipse (4.1cm and 2.65cm);
\node[right] at (4.25,0) {$r_s(l)$};

\begin{scope}
\clip (0,0) ellipse (4.1cm and 2.65cm);
\fill[ergocolor] (0,0) ellipse (2.3cm and 1.45cm);
\end{scope}

\node at (0,2.0) {\small Ergoregion};
\node at (0,1.55) {\small $g_{tt}>0$};

\draw[->, thick] (-5.4,0.35) -- (-2.9,0.2);
\node[left] at (-5.4,0.35) {$E_0>0$};

\fill (-2.6,0.2) circle (2pt);
\node[above] at (-2.6,0.27) {\small decay};

\draw[->, thick] (-2.6,0.2) -- (-1.5,0.05);
\node[below] at (-1.55,0.05) {$E_1<0$};

\draw[->, thick] (-2.6,0.2) -- (-4.6,1.55);
\node[left] at (-4.6,1.55) {$E_2>E_0$};

\draw[->, ultra thick, omegaColor] (2.3,0.22) arc[start angle=5,end angle=35,x radius=2.5,y radius=1.45];
\node[omegaColor, right] at (2.05,0.7) {$\Omega_h(l)$};

\draw[->, thick] (3.35,2.25) -- (3.95,2.65);
\node[right] at (3.95,2.65) {\small increasing $l$};
\end{tikzpicture}
\caption{Penrose energy extraction in the Bumblebee Kerr-Newman-AdS spacetime. The Lorentz-violating parameter $l$ shifts the static limit surface from $r_s^{(0)}$ to $r_s(l)$, enlarging the ergoregion defined by $r_h<r<r_s(l)$. A particle entering the ergoregion decays into a negative-energy fragment that falls through the event horizon and a fragment that escapes to infinity with energy exceeding that of the initial particle. The horizon angular velocity $\Omega_h(l)$ (red arrow) follows the horizon curvature near the event horizon, indicating the direction of frame dragging.}
\label{fig:penrose_LV}
\end{figure}
Figure~\ref{fig:penrose_LV} provides a geometric interpretation of the Penrose process in the presence of Lorentz symmetry breaking. The outward displacement of the static limit surface induced by the Bumblebee parameter $l$ increases the spatial extent of the ergoregion, enlarging the region where the timelike Killing vector becomes spacelike. As a result, the range of admissible negative-energy trajectories is extended, allowing infalling fragments to reach lower conserved energies before crossing the horizon. Simultaneously, the increase in the horizon angular velocity $\Omega_h(l)$ strengthens frame dragging near the event horizon, further decreasing the minimum energy absorbed by the black hole. These combined geometric and kinematic effects lead to a systematic enhancement of the Penrose energy extraction efficiency relative to the Lorentz-invariant Kerr-Newman-AdS spacetime.

\subsection{The proper radial thickness}\label{subsec:PRT}

To quantify the extent of the ergosphere outward from the horizon, we consider its radial thickness as a function of polar angle. The proper radial thickness of the ergosphere at polar angle $\theta$ is defined as:

\begin{equation}
\ell(\theta) =\int _{r_1}^{r_2}\sqrt{g_{rr}(r,\theta)}dr= \int_{r_h}^{r_s(\theta)}\sqrt{\frac{\Sigma(r,\theta)}{\Delta_r(r)}}dr,
\end{equation}
where $r_h$ and $r_s(\theta)$ are the event horizon and static-limit radii, respectively.  
Due to the quartic structure of $\Delta_r$, this integral can be expressed only in terms of elliptic functions and therefore does not yield a compact closed form.

For physical transparency, we therefore adopt a near-horizon, slow-rotation expansion, which leads to
\begin{equation}
	\begin{split}
	\ell(\theta) &\simeq 2a|\sin\theta|\frac{\sqrt{r_h^2+a^2\cos^2\theta}}{ \Delta'_r(r_h)}\\
	&\simeq2a|\sin\theta|\frac{\sqrt{r_h^2+a^2\cos^2\theta}}{ -2 M+2 r_h+\frac{2 l r_h}{1-l}-\frac{2 \Lambda  r_h^3}{3 (1-l)}-\frac{2 \Lambda  r _h\left(a^2+r_h^2\right)}{3 (1-l)}},
	\end{split}
\end{equation}
clearly showing the angular dependence and the linear growth of the ergoregion thickness with the spin parameter.	

\begin{figure}[t]
	\centering
	
	\includegraphics[width=.40\textwidth]{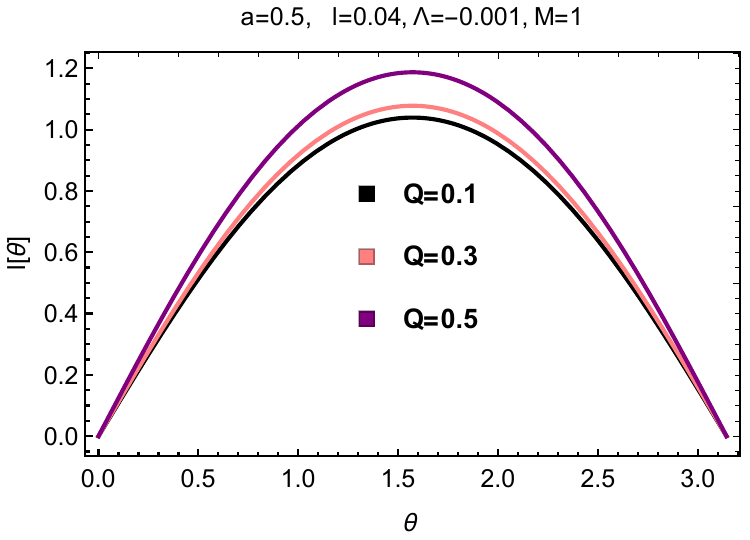}
	
	\caption{Proper thickness as a function of $\theta$ for different values}
	\label{f-14}
\end{figure}

In Fig. \ref{f-14}, we plot the proper thickness as a function of $\theta$ for different values of the charge, while the other parameters are fixed at $\Lambda=-0.001$, $l=0.04$ and $a=0.5$. The proper thickness vanishes at $\theta=0$ and $\theta=\pi$, and reaches its maximum at $\theta=\pi/2$. Moreover, we observe that increasing the charge leads to an increase in the proper thickness for all $0<\theta<\pi$.
\section{Conclusion}\label{sec:conc}

In this work, we have performed a comprehensive analysis of the Bumblebee Kerr-Newman-AdS black hole, a Lorentz-violating extension of the Kerr-Newman-AdS geometry incorporating a dynamical bumblebee vector field. The dimensionless Lorentz-violating parameter $l$ modifies the spacetime curvature, alters both the gravitational and electromagnetic sectors, and induces observable deviations from the predictions of Lorentz-invariant general relativity in strong-field regimes.

The structure of the horizons, determined by the roots of the radial function $\Delta_r$, exhibits strong sensitivity to the black hole parameters $a$, $Q$, $\Lambda$, and $l$. An increase in the Bumblebee parameter $l$ enlarges the horizon radius and shifts the extremality condition, leading to modifications in the causal structure of the spacetime. The quartic nature of the horizon equation allows for non-extremal, extremal, and naked singular configurations, demonstrating that Lorentz symmetry breaking directly influences both the geometric and stability properties of the black hole.

Black hole thermodynamics is significantly affected by Lorentz violation. Surface gravity calculations and horizon geometry reveal that the Hawking temperature decreases with increasing horizon size, while rotation and Lorentz symmetry breaking shift the critical points associated with evaporation. The derived expressions for the remnant mass and radius show that rotational effects enhance stability by preventing total evaporation. Entropy and heat capacity exhibit divergences corresponding to phase transitions between stable and unstable regimes, with the presence of $l$ extending the domain of thermodynamic stability and supporting the existence of long-lived black hole remnants.

The sparsity of Hawking radiation, quantified by the parameter $\eta$, was analyzed to describe emission properties. The inverse proportionality between $\eta$ and the Hawking temperature implies that Lorentz violation produces more discrete emission for small black holes. As entropy increases, $\eta$ decreases, signaling the transition to nearly continuous radiation. The Bumblebee parameter further reduces the radiation flux and prolongs the lifetime of evaporating black holes, introducing measurable modifications to the emission spectrum that could serve as observational signatures of Lorentz symmetry breaking.

Photon trajectories and the optical properties of the black hole were investigated via the Hamilton--Jacobi formalism. The derived impact parameters determine the photon sphere and the celestial coordinates defining the apparent shadow. Rotation induces frame-dragging effects and D-shaped shadow deformation, while electric charge reduces the shadow radius. The Lorentz-violating parameter $l$ introduces asymmetry and horizontal elongation, and the cosmological constant modifies the focusing of light rays. Quantities such as the shadow area, distortion, and angular diameter vary systematically with the model parameters, indicating that even small Lorentz-violating effects produce detectable deviations in shadow morphology.

The ergoregion, defined by the event horizon and static limit surface, expands and deforms with increasing rotation and Lorentz violation. Its volume depends on $a$, $l$, $Q$, and $\Lambda$, while the efficiency of the Penrose process increases in the presence of Lorentz symmetry breaking. The frame-dragging angular velocity $\Omega=-g_{t\phi}/g_{\phi\phi}$ and the horizon angular velocity $\Omega_h(l)$ both increase with the Lorentz-violating parameter $l$, as shown in Fig.~\ref{fig:penrose_LV}. This enhancement corresponds to a stronger rotational dragging of spacetime near the event horizon, which lowers the threshold for negative-energy infall. Consequently, the fragment absorbed by the black hole can attain a more negative energy $E_1$, while the escaping fragment carries a correspondingly larger energy $E_2>E_0$. The Penrose efficiency $\eta=-E_1/E_0$ therefore reaches higher maximal values $\eta_{\max}$ compared to the Lorentz-invariant case. Physically, Lorentz-symmetry breaking strengthens the interaction between the black hole's rotation and particle dynamics: the enlarged ergoregion increases the spatial domain available for energy extraction, and the intensified frame dragging near the horizon allows infalling fragments to achieve deeper negative-energy states. These combined geometric and kinematic effects lead to a significant enhancement of rotational energy extraction efficiency, demonstrating that deviations from exact Lorentz invariance in the gravitational sector can directly influence the energetics of processes around rotating black holes.

Taken together, the Bumblebee Kerr-Newman-AdS black hole exhibits profound modifications across geometric, thermodynamic, radiative, and optical domains due to Lorentz symmetry breaking. The parameter $l$ systematically influences the horizon structure, modifies black hole evaporation properties, alters the sparsity and flux of Hawking radiation, reshapes photon trajectories, and enlarges the ergoregion, thereby increasing the potential for rotational energy extraction. These results provide a robust theoretical framework for investigating Lorentz-violating effects in strong gravitational fields and offer predictive guidance for observational tests using high-resolution black hole imaging, precise measurements of emission spectra, and gravitational wave observations.

The Bumblebee Kerr-Newman-AdS spacetime thus constitutes a precise model for connecting modified gravity theories with observationally accessible phenomena. Even minimal deviations from Lorentz invariance (\(l \in [0,1)\)) produce significant changes in horizon geometry, radiation emission, shadow structure, and energy extraction efficiency. Future research could explore quasinormal mode spectra, accretion-disk radiation signatures, gravitational-wave echoes in Lorentz-violating scenarios, and holographic correspondences within AdS/CFT frameworks. Such investigations have the potential to reveal the observational imprint of spontaneous Lorentz symmetry breaking and to deepen our understanding of spacetime structure in extreme gravitational environments.

\section*{Acknowledgement}
 H.H is grateful to Excellence project FoS UHK 2203/2025-2026 for the financial support. 









\end{document}